\documentclass[twocolumn,showkeys,prd,nofootinbib,floatfix,preprintnumbers,superscriptaddress]{revtex4-1}

\usepackage{multirow}
\usepackage{array}
\usepackage{hyperref}
\usepackage[normalem]{ulem}
\usepackage[utf8]{inputenc}
\usepackage{amsfonts,amsmath,amssymb} 
\usepackage{graphicx,graphics,color}
\usepackage{xcolor} 
\usepackage{gensymb}
\usepackage{longtable}
\usepackage{bbding}
\usepackage{subfigure}
\usepackage{multirow}
\usepackage{float}
\usepackage{ulem}
\usepackage{soul}

\begin{document}

\title{Exploring Models of Running Vacuum Energy with Viscous Dark Matter\\ from a Dynamical System Perspective}

\author{Normal Cruz}
\email{norman.cruz@usach.cl}
\affiliation{Departamento de F\'isica, Universidad de Santiago de Chile,\\Avenida V\'ictor Jara 3493, Estaci\'on Central, 9170124 Santiago, Chile}

\author{Gabriel G\'omez}
\email{gabriel.gomez.d@usach.cl}
\affiliation{Departamento de F\'isica, Universidad de Santiago de Chile,\\Avenida V\'ictor Jara 3493, Estaci\'on Central, 9170124 Santiago, Chile}

\author{Esteban Gonz\'alez}
\email{esteban.gonzalez@uac.cl}
\affiliation{Direcci\'on de Investigaci\'on y Postgrado, Universidad de Aconcagua, Pedro de Villagra 2265, Vitacura, 7630367 Santiago, Chile.}

\author{Guillermo Palma}
\email{guillermo.palma@usach.cl}
\affiliation{Departamento de F\'isica, Universidad de Santiago de Chile,\\Avenida V\'ictor Jara 3493, Estaci\'on Central, 9170124 Santiago, Chile}

\author{\'Angel Rinc\'on}
\email{angel.rincon.r@usach.cl}
\affiliation{Departamento de F\'isica, Universidad de Santiago de Chile,\\Avenida V\'ictor Jara 3493, Estaci\'on Central, 9170124 Santiago, Chile}

\begin{abstract}
Running vacuum models and viscous dark matter scenarios beyond perfect fluid idealization are two appealing theoretical strategies that have been separately studied as alternatives to solve some problems rooted in the $\Lambda$CDM cosmological model. In this paper, we combine these two notions in a single cosmological setting and investigate their cosmological implications, paying particular attention in the interplay between these two constituents in different cosmological periods. Specifically, we consider a well-studied running vacuum model inspired by renormalization group, and a recently proposed general parameterization for the bulk viscosity $\xi$.
By employing dynamical system analysis, we explore the physical aspects of the new phase space that emerges from the combined models and derive stability conditions that ensure complete cosmological dynamics. We identify four distinct classes of models and find that the critical points of the phase space are non-trivially renewed compared to the single scenarios. We then proceed, in a joint and complementary way to the dynamical system analysis, with a detailed numerical exploration to quantify the impact of both the running parameter and the bulk viscosity coefficient on the cosmological evolution. Thus, for some values of the model parameters, numerical solutions show qualitative differences from the $\Lambda$CDM model, which is phenomenologically appealing in light of cosmological observations.

\end{abstract}

\maketitle
\section{Introduction}

The standard cosmological model, also known as the $\Lambda$CDM model, is currently the most successful theoretical framework for describing the evolution of the universe \cite{Peebles:1984ge,Peebles:1994xt,Krauss:1995yb,Ostriker:1995su}. However, as the precision of cosmological observations is continuously increasing \cite{Alam:2016hwk,Planck:2018vyg,Jones:2017udy,DES:2021wwk,SNLS:2010pgl,SupernovaCosmologyProject:2011ycw,SNLS:2011lii,ACT:2020gnv,2dFGRS:2005yhx,SDSS:2006lmn,BOSS:2016wmc,HSC:2018mrq,Heymans:2020gsg,HSC:2018mrq}, the model is facing new challenges in maintaining observational consistency \cite{Bullock:2017xww,Freedman:2017yms}. Despite significant progress in our current understanding of the universe, still, we have some issues that require further investigation. These include: i) the Hubble and $\sigma_{8}$ tensions, which refer to discrepancies between the values predicted by the model and observations of the Hubble constant \cite{Freedman:2017yms,Verde:2019ivm} and the amplitude of matter fluctuations on large scales \cite{Macaulay:2013swa,Battye:2014qga,Alam:2016hwk,Abbott:2017wau}, respectively; and ii) a lack of comprehension of the physics involved in the dark sector. 
One of the most crucial conceptual problems is related to the nature of dark matter (DM) \cite{Arbey:2021gdg}, which comprises approximately 80$\%$ of the total matter of the universe.  Another well-known problem associated with the $\Lambda$CDM model is the cosmological constant problem \cite{1995AmJPh63620A,Hobson:2006se}(CC problem for short), which arises due to the discrepancy between the estimated value for the vacuum energy density (VED) provided by quantum field theory and the observed value inferred by type Ia supernovae (SNe Ia) \cite{SupernovaSearchTeam:1998fmf}. 

 Given the significant discrepancies previously mentioned, as well as the physical argument that an expanding universe is not expected to have a static vacuum energy density, scientists have suggested to explore a smooth time-dependence of it. One option along this line of thinking is to use a decreasing function for the cosmological constant that could potentially address not only the Hubble constant tension but also bring the predicted value closer to the observed one \cite{Fritzsch:2016ewd}. A more general time-dependence of the vacuum energy density has been shown to be implicitly given through the Hubble constant and its time derivatives $\rho_{\text{vac}}(H,\dot{H})$ \cite{Sola:2015rra}, which is motivated by perturbative results of Quantum Field Theory in a curved classical background \cite{Parker:2009uva}. This is an interesting cosmological scenario in contrast to other physical proposals based on dynamical dark energy \cite{Amendola:2015ksp,DiValentino:2017gzb}, which assume that the cosmological constant is small or negligible compared to the total energy density \cite{Amendola:2015ksp}. 

A more generalized Ansatz for the vacuum energy density has been proposed by considering Renormalization Group ideas, which consider it as a running quantity depending on the typical energy-scale of the processes involved \cite{Shapiro:2000dz}. This strategy has been used in ref. \cite{Sola:2013gha} to deduce a functional form for the vacuum energy density, which depends dynamically on the Hubble constant and on its time derivative. We will use this suited Ansatz for $\rho_{\text{vac}}$ to describe the late universe evolution, which replaces a constant vacuum energy density with its ``running" counterpart. For technical details see \cite{Sola:2015rra}. 

The other dark component of the universe, which is commonly described by a pressureless fluid, is known as cold dark matter (CDM). This is responsible for the structure formation of the Universe. In a wider physical ambit DM can include viscosity and even warmness due to late decoupling from the primordial plasma.  In fact, some of the present tensions of the standard model have been alleviated with the inclusion of viscosity in the dark sector. For example, the Hubble tension, which exhibits a discrepancy of $4.4\sigma$ between the measurements obtained from Planck CMB and the locally ones obtained in \cite{Riess:2019cxk} for $H_{0}$, has been tackled in \cite{tensionH0,Wilson:2006gf,BulktensionH}.  The $\sigma_{8}$ tension (where $\sigma_{8}$ is the r.m.s. fluctuations of perturbations at $8h^{-1}$~Mpc scale) that emerges when confronting large-scale structure (LSS) observations and Planck CMB data \cite{remedyforplankanlssdata,sigma8}, can be attenuated assuming a viscous DM component \cite{remedyforplankanlssdata}. The EDGES experiment has observed an excess of radiation at $z\approx17$ 
\cite{Bowman:2018yin},
which is not predicted by the $\Lambda$CDM model during the reionization epoch. This excess can be indeed explained by the presence of a viscosity DM component \cite{linia21cm}.

Despite the fact that dynamical viscous dark matter models and the presence of a running vacuum energy density have the potential to alleviate some of the tensions present in the $\Lambda$CDM model, it should be noted that they have, separately, limitations in successfully describing the entire cosmological evolution, besides the fundamental physical conceptions.
However, by combining both hypotheses, we seek for a more complete physical scenario to address the limitations of each single approach, aiming to associate, for instance, the rate of structure formation to dissipative effects of CDM, while the tension related to the Hubble constant to the running vacuum energy density. The primary question we want to address in this paper is whether incorporating these two ideas into a more comprehensive cosmological framework could provide a completely consistent cosmological evolution.

Moreover, together with the above-mentioned argument, in ref. \cite{Ashoorioon:2023jwf} different varying viscous DM models are used to address the Hubble and $\sigma_{8}$ tensions of the standard $\Lambda$CDM model. It is shown that although the proposed dissipative models tend to reduce the $\sigma_{8}$ tension, they aggravate the Hubble tension, which leads the authors to conclude that in addition to DM viscosity a dynamical presence of relativistic universe components or dark energy should be required to simultaneously alleviate both tensions.

In light of the aforementioned discussions concerning the two possible modifications of the  dark sector of the standard model, the main goal of the present article is to investigate -using the dynamical system approach combined with numerical methods- the critical points and their stability properties, which account for the dynamical expansion of the universe. A fundamental requirement for a cosmological model is to accurately describe the complete evolution of the universe, which includes radiation, matter, and dark energy periods. Although the radiation-dominated period is often neglected for simplicity, it is not trivial to incorporate it into viscous dark matter models using certain parametrizations, as shown in \cite{Gomez:2022qcu}. Therefore, it is necessary to ensure that the three main eras of cosmic evolution are present as critical points with the appropriate stability properties within a consistent model parameter region. The proposed model includes a running vacuum density and dissipative dark matter. 

Lastly, a previous advance obtained by using dynamical system analysis, which considers also a running vacuum and viscous DM with the particular parameterization $\xi=\xi_{0}H$, where $\xi$ is the usual bulk viscosity coefficient, was performed in \cite{N:2022rbh}. Nevertheless, aside of the particular Ansatz for the dissipation used, the analysis is restricted to late times, as the radiation component was not included. In the present paper, we aim to address these limitations by considering a more general parametrization for the bulk viscosity given by \cite{Gomez:2022qcu} and a running vacuum energy density. This general parametrization has the advantage of  including simultaneously kinematic effects represented by the Hubble constant, and dynamical ones played by the dark matter energy density. In this way, the bulk viscosity associated to DM is consistently handled vanishing as matter density does. More relevant is the fact that Friedmann's equations can be written in the form of an autonomous dynamical system for any value of the exponent that characterizes dissipation within Eckart’s framework of relativistic non-perfect fluids. In addition, we incorporate the radiation component of the universe for consistency, as discussed above. 
 
The present paper is organized as follows:
In Section \ref{sec:Model} the main ingredients of the model, the evolution equations, and the physical motivations behind the Ans\"atze chosen for the running vacuum energy density and the DM viscosity coefficient are present. 
In Section \ref{sec:Dynamical} a dynamical system analysis is performed, whose space is spanned by the phase variables $\Omega_r$, $\Omega_m$ and $\Omega_{\rm vac}$ (see Eq.\ref{sec3:eqn3} for details). Furthermore, the fixed points and their stability properties are explicitly computed. In particular, based on the parametrization of the running vacuum density and the corresponding one for the dissipative DM, four prominent models are throughout studied in subsections \ref{sec:Model1Dynamical}-\ref{sec:Model4Dynamical}.
In Section \ref{sec:Numerical} we describe the numerical procedure used for the integration of the suited classes of models considered in this paper. 
Finally, in Section \ref{sec:Remarks} the main findings of the present article are summarized, and a physical discussion of the cosmological scenarios resulting from the models studied is provided.

\section{\label{sec:Model}The model}
In this section we will describe the cosmological model we propose to study the cosmological dynamics of the universe. It represents a two-fold extension of the $\Lambda$CDM model, which includes two essentials aspects that has been extensively used to alleviate or solve some problems associated to the standard cosmological model. Firstly, we consider a vacuum energy density described by a running coupling depending on both the Hubble parameter and its cosmological time derivative $\rho_{\rm vac}(H,\dot{H}) $. This Ansatz is not only motivated by the intuitive observation that an expanding universe quite improbably would preserve a static value throughout its complete evolution, but it is also motivated by fundamental physics, in fact, a smooth evolving vacuum energy density is suggested by quantum field theory in curved spacetime (see \cite{Sola:2015rra} and references therein). Secondly, we propose a more realistic fluid description of the dark matter component including dissipation through a bulk viscosity coefficient, which has been recently proposed by the authors \cite{Gomez:2022qcu}. This proposal leads to remarkable advantages, such that the cosmological time evolution equations can be written in a form of an autonomous dynamical system suited to be studied by the stability theory, and that the bulk viscosity effects fade away when the dark matter density vanishes. 

As already mentioned in the above paragraph, for the running vacuum vacuum energy density we will use a two-parameter model inspired by a phenomenological application of Renormalization Group analysis, whose cosmological consequences has been studied in \cite{Sola:2015rra}, which can be written as  

\begin{equation}
    \rho_{\rm vac}(H)=\frac{3}{8\pi G_{N}} 
    \Bigl(
    c_{0}+\nu H^{2}+\tilde{\nu}\dot{H}
    \Bigl)
    +
    \mathcal{O}(H^{4}),
    \label{sec2:eqn1}
\end{equation} 
where $\nu$ and $\tilde{\nu}$, are both dimensionless parameters, and it is expected that $|\nu|$ and $|\tilde{\nu}|$ should be lower than one. Indeed, according to QFT calculations, the more suitable values of the set $\{ \nu, \tilde{\nu} \}$ rounding $10^{-3}$, which also it has been obtained from  
the constraints using SNe Ia+BAO+H(z)+LSS+CMB cosmological observations \cite{peracaula2018dynamical}.  The above phenomenological Ansatz is suited for the wide range of the universe expansion excluding early times, as the Hubble parameter grows very fast when for instance, the inflationary epoch is approached.

We are particularly interested in two sub-lasses of running vacuum models for checking their cosmological viability by dynamical system perspective, directly based on Eq. \eqref{sec2:eqn1}.
The constant parameters $\nu$ and $\tilde{\nu}$ account for the dynamical character of the vacuum energy density, $c_{0}$ is a constant determined by the boundary condition $\rho_{\rm vac}(H_{0},\dot{H}^{(0)})=\rho_{\rm vac}^{(0)}$, where the superscript $(0)$ refers to the present value, i.e. $a_0=1$. Thus, by exploiting the independence of both dynamical contributions, two classes of running vacuum models can arise. 
The first possibility we want to investigate corresponds to the case with $\tilde{\nu}=0$. This is indeed one of most treated cases in scenarios of variable vacuum energy density. As to the second class of running vacuum models, we will focus on the particular choice $\tilde{\nu}=\nu/2$, since it has the potential advantage of alleviating some tensions permeated in the $\Lambda$CDM cosmological model \cite{SolaPeracaula:2021gxi}. The inclusion of such term has appealing consequences in the conservation law for the involved components because, as it will be seen, it allows to write the vacuum energy density only in terms of the matter component in contrast to the first class of models\footnote{Notice that even though $\dot{H}$ involves a term associated to the radiation energy density it cancels out with the coming one from $H^{2}$.}. Hence, when the radiation component is considered as a part of the total energy density for the first class of models, the vacuum energy density will depend on it, whereby all components are coupled directly to each other apart from gravitational interaction. This feature can provide appreciable differences in the background cosmological evolution. It is expected however that the radiation component naturally becomes negligible in the late-time dynamics, and the effects of the running of the vacuum in the radiation era can be considerably small to impact the thermal history of the Universe. But, are there some consequences of the running of the vacuum energy density from the dynamical system perspective on the emerging critical points? If so, how much do the stability conditions are changed with respect to the reference $\Lambda$CDM model? These aspects are the ones to be assessed in this work.

According to the previous discussion and considering the running vacuum model in Eq.~(\ref{sec2:eqn1}), the Friedmann and the acceleration equations are respectively written as
\begin{eqnarray}
    3H^{2}&=&8\pi G_{N} \left(\rho_{r}+\rho_{m}+\rho_{\rm vac}\right),\label{sec2:eqn2}\\ 
    3H^{2}+2\dot{H}&=&-8\pi G_{N} \left(P_{r}+P_{m}^{\rm eff}+P_{\rm vac}\right), \label{sec2:eqn3}
\end{eqnarray}
where the usual polytropic relation for radiation $P_{r}=\rho_{r}/3$ is set, and the one for the vacuum energy density  $P_{\rm vac}=-\rho_{\rm vac}$ holds  provided that Eq.~(\ref{sec2:eqn1}) is identified as the true vacuum energy density. One would expect however some deviation from $w_{\rm vac}=-1$ at, for instance, early times given the dependence of $\rho_{\rm vac}$ on the Hubble parameter: Eq.~(\ref{sec2:eqn1}) tell us that once $\rho_{\rm vac}$ is promoted to a dynamical quantity, it can evolve so that an effective equation of state may take an appreciably different value from the standard one. 

For the dark matter fluid, the bulk viscous pressure $\Pi$ is introduced as an effective pressure to allow more phenomenological outcomes within a cosmological setting beyond the standard running vacuum models: 
\begin{equation}
    P_{m}^{\rm eff}=P_{m}+\Pi=-3H\xi,\label{sec2:eqn4}
\end{equation}
where $\xi$ is the bulk viscosity coefficient that respects the second law of the thermodynamics provided that $\xi>0$. As previously said, this extra ingredient has been introduced to get a more realistic fluid description of DM, and for enriching the phase space of the system and also for suitable comparison with typical bulk viscosity models with $\nu=\tilde{\nu}=0$ which are minimal extensions of the $\Lambda$CDM cosmological model. In doing so, we will use a recently proposed general parameterization for the viscosity coefficient $\xi$ 
\begin{equation}
    \xi=\frac{\xi_{0}}{8\pi G_{N}} H^{1-2s} H_{0}^{2s}\left(\frac{\rho_{m}}{\rho_{m}^{0}} \right)^{s}=\frac{\hat{\xi}_{0}}{8\pi G_{N}} H\; \Omega_{m}^{s},\label{sec2:eqn5}
\end{equation}
(for technical aspects see ref. \cite{Gomez:2022qcu}).
The above parametrization has several advantages, among them we mention that it encompasses the well known models $\xi=\xi(H)$ (corresponding to $s=0$) and $\xi=\xi(\rho_{m})$, or more precisely\footnote{Notice that taking $s=1$ and assuming $H\propto \rho_{m}^{1/2}$ lead also to  $\xi\sim \rho_{m}^{1/2}$ . We stress however that this limit is achieved only when the universe is in the matter domination epoch. So it is expected that $s=1$ and $s=1/2$ are different when dark matter is subdominant but not negligible.} $\xi\sim \rho_{m}^{1/2}$ (for $s=1/2$) and that, in turns, it is very useful when writing the resulting evolution equations in the form of autonomous system through the second equality (r.h.s. of Eq.\eqref{sec2:eqn5}).
Notice that $\hat{\xi}_{0}$ and $\xi_{0}$ are both dimensionless constants related to each other by $\hat{\xi}_{0}=\frac{\xi_{0}}{\Omega_{m}^{0}}$. It is very instructive now to formulate the conservation law for each component for the two classes of models discussed above. The Bianchi identities establish thus the global conservation law
\begin{equation}
    \dot{\rho}_{r} + 4 H \rho_{r} + 
 \dot{\rho}_{m} + 
  3 H (\rho_{m} + \Pi) = -\dot{\rho}_{\rm vac}.\label{sec2:eqn6}
\end{equation}
Let us write the conservation law considering Eqs.~(\ref{sec2:eqn1}), (\ref{sec2:eqn2}) and (\ref{sec2:eqn3}) keeping $\tilde{\nu}$ free to trace its effect at the level of the conservation equations:
%
\begin{align}
\begin{split}
    &\dot{\rho}_{r} + 4 H \rho_{r} + 
 \dot{\rho}_{m} + 
  3 H (\rho_{m} + \Pi) = 
  \\
  & \nu \Bigl(3 (\Pi + \rho_{m}) + 4 \rho_{r}\Bigl) H + \frac{3}{2} \tilde{\nu} \left(\dot{\Pi} +\dot{\rho}_{m} +\frac{4}{3} \dot{\rho}_{r}\right).\label{sec2:eqn7}
\end{split}
\end{align}
%
Considering, for the time being, the lineal relation $\tilde{\nu} = A\nu$, with $A$ being some arbitrary value. So once terms associated to the same nature's fluid have been grouped, the continuity equations for each fluid take the form
\begin{align}
   &\dot{\rho}_{r}(1-2A\nu )+4H\rho_{r}(1-\nu) =0,\label{sec2:eqn8}
   \\
   &\dot{\rho}_{m}\left(1-\frac{3}{2}A\nu\right)+3H(\rho_{m}+\Pi)(1-\nu)-\frac{3}{2}A\nu\dot{\Pi} =0,\label{sec2:eqn9}
\end{align}
where $\dot{\Pi}$ is a function of $\rho_{m}$ and $\dot{\rho}_{m}$ determined by the viscous model of Eq.~(\ref{sec2:eqn5}). It is interesting to see that for the particular value $A=1/2$, or equivalently $\tilde{\nu}=\nu/2$, the standard conservation equation for radiation holds whereas the one for the dark matter fluid is modified\footnote{Notice that there exists a value for $A$ ($A=2/3$), along with turning off the bulk viscosity, for which the dark matter fluid follows the standard form. We are however interested in the model with $A=1/2$ without prejudice against other values that may lead to interesting phenomenological features in the radiation era.}. It means that necessarily one of those conservation equations must be modified at the cost of allowing the running of the vacuum energy in the form given by Eq.~(\ref{sec2:eqn1}). As to the evolution equation for the vacuum energy density, it has two possible contributions according to the right hand side of Eq.~(\ref{sec2:eqn7}). For the first class of models (with $\tilde{\nu}=0$), we can see that  the energy densities of the fluids, and not their derivatives, will contribute to the evolution of $\rho_{\rm vac}$. This implies that the prefactor $(1-\nu)$ in the continuity equations (Eqs.~(\ref{sec2:eqn8}) and (\ref{sec2:eqn9})) can not be cancel out unless one goes to the trivial case $\nu=0$. By the contrary, turning on $\tilde{\nu}$ implies that the evolution equations for radiation and dark matter, in addition to the one of the viscous pressure, must be considered to account properly for the time evolving vacuum energy density. This is, in fact, the main difference between both classes of models we want to investigate along with the possibility of taking different values of the power $s$ in the bulk viscosity coefficient of Eq.~(\ref{sec2:eqn5}). This is specified in Table \ref{Models}. 
Having thus specified the main ingredients of the general model, we will proceed to perform dynamical system analysis in the next section.

\begin{table*}[htp]
\centering  
\caption{Classification of the viscous running cosmological models to be studied according to the dependence of $\rho_{\rm vac}$ either on $H$ only (\textit{first class} $\tilde{\nu}=0$ and $s = 1/2$) or on both $H$ and $\dot{H}$ (\textit{second class models}  $\tilde{\nu}=\nu/2$ and $s = 1, 1/2, 0$ respectively).}
\begin{ruledtabular}
\begin{tabular}{cccc}
Label & Class of model & Bulk viscosity exponent $s$ &
  \\ \hline
 $\text{Model 1}$ & $\text{First class}\;\tilde{\nu}=0 $ & $1/2$\\
  $\text{Model 2}$ & $\text{Second class}\; \tilde{\nu}=\nu/2$ & $1$\\
   $\text{Model 3}$ & $\text{Second class}\;\tilde{\nu}=\nu/2$ & $1/2$\\
    $\text{Model 4}$ & $\text{Second class}\;\tilde{\nu}=\nu/2$ & $0$\\
\end{tabular}
\end{ruledtabular}\label{Models}
\end{table*}
%

\section{\label{sec:Dynamical} Dynamical system analysis}
We start by defining the dimensionless variables that span the phase space of the system and allows us to rewrite their dynamics in the form of an autonomous system. For practicality, such variables are chosen essentially to describe the density parameters associated to each fluid 
\begin{align}
\begin{split}
\Omega_{r} &\equiv \frac{8\pi G_{N} \rho_{r}}{3H^{2}},
\\
\Omega_{m} &\equiv \frac{8\pi G_{N} \rho_{m}}{3H^{2}}, 
\\
\Omega_{\rm vac} &\equiv \frac{8\pi G_{N} \rho_{\rm vac}}{3H^{2}}.
\label{sec3:eqn1}
\end{split}
\end{align}
Therefore, the Friedmann constraint takes the usual form
\begin{equation}
\Omega_{r}+\Omega_{m}+\Omega_{\rm vac}=1,\label{sec3:eqn2}
\end{equation}
and the evolution equations for radiation, dark matter and the vacuum energy are, respectively, for the first class of models

\begin{widetext}
\begin{align}
\begin{split}
\Omega_{r}^\prime&=\Omega_{r} (-1 + 4 \nu - 3 \hat{\xi}_{0} \Omega_{m}^{s} + \Omega_{r} - 3 \Omega_{\rm vac}) , 
\\
\Omega_{m}^\prime&= -3 (-1 + \nu) \hat{\xi}_{0} \Omega_{m}^{s} - 3 \hat{\xi}_{0} \Omega_{m}^{1 + s} + \Omega_{m} (3 \nu + \Omega_{r} - 3 \Omega_{\rm vac}), 
\\
\Omega_{\rm vac}^\prime&=
-3 \nu \Omega_{m} + 3 \hat{\xi}_{0} \Omega_{m}^{s} (\nu - \Omega_{\rm vac}) - 
 3 (-1 + \Omega_{\rm vac}) \Omega_{\rm vac} + \Omega_{r} (-4 \nu + \Omega_{\rm vac}), \label{sec3:eqn3}
\end{split}
\end{align}
\end{widetext}

and for the second class as follows
\begin{widetext}
\begin{align}
\begin{split}
\Omega_{r}^\prime&=\Omega_{r} (-1 - 3 \hat{\xi}_{0} \Omega_{m}^{s} + \Omega_{r} - 3 \Omega_{\rm vac}) , 
\\
\Omega_{m}^\prime&=-\frac{\Omega_{m} \left(
    -12 (-1 + \nu) \hat{\xi}_{0} \Omega_{m}^s 
    - 9 \nu \hat{\xi}_{0}^{2} \Omega_{m}^{2s} + 
   6 (-2 + 3 \nu) \hat{\xi}_{0} \Omega_{m}^{1 + s} + 
   \Omega_{m} ((4 - 3 \nu)\Omega_{r} + 
   3 (\nu + (-4 + 3 \nu) \Omega_{\rm vac}))\right)
   }{(-4 + 3 \nu) \Omega_{m} - 3 s \nu\hat{\xi}_{0} \Omega_{m}^{s}},
\\
\Omega_{\rm vac}^\prime&= - \nu (3 \Omega_{m} - 3\hat{\xi}_{0}\Omega_{m}^{s} + 4 \Omega_{r}) + \frac{1}{4}(-3 + 
      3 \hat{\xi}_{0} \Omega_{m}^{s} - \Omega_{r} + 3 \Omega_{\rm vac}) (-3 \nu \Omega_{m} + 
      3 \nu \hat{\xi}_{0} \Omega_{m}^{s} - 
      4 (\nu \Omega_{r} + \Omega_{\rm vac})) +  
\\ 
& \ \ \ \  \frac{\nu}{4} \left((-3 + 3 s \hat{\xi}_{0} \Omega_{m}^{-1 + s}) \Omega_{m}^\prime - 
      4 \Omega_{r}^\prime\right). \label{sec3:eqn4}
\end{split}
\end{align}
\end{widetext}
Here the prime denotes derivative with respect to $ \lambda \equiv \ln a$. It is worthwhile to emphasizing that the evolution equation for radiation has been included here for illustration purposes and to write in a compact form the evolution equation for the vacuum energy density parameter since the system can be reduced to two dimensional phase space with the help of Eq.~(\ref{sec3:eqn2}). As discussed, for the second class of models the evolution equation for the radiation density parameter holds the standard form in the absence of bulk viscosity as in the $\Lambda$CDM model, and the evolution equation for the vacuum energy density parameter involves derivatives of the other components as can be evidenced in the last line of Eqs.~(\ref{sec3:eqn4}). 

The effective equation of state parameter is defined as usual
\begin{equation}
w_{\rm eff}=-\frac{2}{3}\frac{H^\prime}{H}-1, \label{barotropic_const}
\end{equation}
with
\begin{equation}
\frac{H^\prime}{H}=\frac{1}{2} (-3 + 3 \hat{\xi}_{0}  \Omega_{m}^{s} - \Omega_{r} + 3  \Omega_{\rm vac}),\label{sec3:eqn5}
\end{equation}
enclosing however the main features discussed above since the energy density parameters, at the critical points, depend on the model parameters $\nu$ and $\hat{\xi}_{0}$, as we shall see. Though so far we have been attempting to describe in a general way the effects of the viscosity associated to the dark matter fluid, it is necessary to take at this point some specific values of the power $s$ in Eq.~(\ref{sec2:eqn5}) for both classes of models in order to carry out suitably the phase space analysis. We will refer henceforth to model 1 for the fist class of model (Eq.~(\ref{sec3:eqn3})) with $s=1/2$ while, for the second class of models (Eq.~(\ref{sec3:eqn4})), model 2 with $s=1$, model 3 with $s=1/2$ and model 4 with $s=0$. In this regards the free parameters of the models are $\nu$ and $\hat{\xi}_{0}$.

It is expected however that these parameters be very small $\nu,\hat{\xi}_{0}\ll 1$, by construction of the theory itself  in addition to the thermodynamical argument (i.e. $\hat{\xi}_{0}>0$), to account properly for the late-time background dynamics.  Different observational constraints suggest also that there parameters are very small. Though those estimations are not strictly applicable for the present models, they will serve as a reference value in the dynamical analysis. Nevertheless, for the sake of generality of our analysis we will take $\nu$ free to see what kind of restriction we can infer from the dynamical system analysis unless stated otherwise. Moreover, we will discard \textit{a priori} any possible solution for which those parameters break the aforementioned  conditions. General stability conditions however will be shown for a better comprehension of how the sign works for determining the dynamical character of the critical point. When one of those free parameters is kept fixed it means that the phase space is insensitive to it. So we will be left with just one parameter to investigate changes in the general stability conditions for the critical points.

\subsection{\label{sec:Model1Dynamical} Model 1: $\tilde{\nu}=0$ and $s=1/2$}
Taking $s=1/2$ in the system Eq.~(\ref{sec3:eqn3}) a  set of 5 critical points is found and reported in Table \ref{M1:critical points}. The first critical point listed below describes non-standard radiation ($\rm Ia$) because of the presence of the parameter $\nu$ or a most general critical point ($\rm Id$) depending on both $\nu$ and $\hat{\xi}_{0}$. Notice also that there may be apparently a sort of degeneracy between $\nu$ and $\hat{\xi}_{0}$ in the critical point ($\rm Id$), however the equation of state is only sensitive to $\nu$. These results are fully consistent with our preliminary expectations regarding the first class of models. 

The critical point  ($\rm Ib$) accounts for dark matter domination and exhibits some trace of the running effects of the vacuum energy density as well as the effects of the bulk viscosity but, for the latter, in an effective way in the equation of state parameter. An interesting feature of this point is that if $\hat{\xi}_{0}\sim \mathcal{O}(1)$, what we will refer to as strong viscous regime, it can drive the current acceleration expansion of the universe through the bulk viscosity effect when the negative branch is considered. Accordingly, for $\hat{\xi}_{0}$ positive and the range $1-\hat{\xi}_{0}^2<\nu<1$, it can reveal a phantom-like behavior, otherwise it will be  $w_{\rm eff}>-1$ for the range $\nu<1-\hat{\xi}_{0}^2$. Interestingly, stability conditions are compatible only with the parameter space associated to a phantom-like solution (see later a more detailed discussion about stability).  

The point ($\rm Ic$) is a sort of scaling solution on account of the parameter $\hat{\xi}_{0}$ describing de-Sitter-like accelerated expansion in the same fashion as the standard critical point ($\rm Ie$) does. From here it is conclusive to say that the effect of the running vacuum energy density is surprisingly not present in the critical points that can potentially drive the current acceleration\footnote{Despite its incapability of addressing the accelerated expansion all on its own, the running vacuum energy density may determine the phantom-like character in the case of an unified fluid scenario ($\hat{\xi}_{0}\sim \mathcal{O}(1)$) as an alternative solution to the de-Sitter solution through the critical point ($\rm Ib$). Here the bulk viscosity dark matter fluid and the running vacuum energy density can be seen as unified fluid description of dark sector.}. It does not mean however that late-time measurements of the background cosmology are completely insensitive to the running effect as was recently assessed in \cite{SolaPeracaula:2017esw}.   Although this model is not so different from the standard $\Lambda$CDM cosmological model under the scrutiny of parameter estimation, this can fit (slightly) better the background data. Nevertheless, high-redshift measurements can also provide significant evidence of the running vacuum energy density effect to see any deviation from the $\Lambda$CDM model \cite{SolaPeracaula:2021gxi}.

Moreover, the running effects can play a crucial role at the perturbation level, leading possibly to different conclusions about structure formation compared to the $\Lambda$CDM model \cite{SolaPeracaula:2017esw}. With the inclusion of bulk viscosity, a richer scenario is expected not only at the background level but also to account for the matter density perturbations. This is an open question that should be dealt with in the future. 

Two interesting sub-manifolds of this model are achieved when the bulk viscosity is turned off ($\hat{\xi}_{0}=0$) and when the limit $\nu=1$ is taken. The latter possibility leads clearly to the $\Lambda$CDM model. This is however a direct consequence of taking $\tilde{\nu}=0$ and it is by no means the formal way of recovering the $\Lambda$CDM model in the more general setup. Finally notice that the existence of the critical points and the request of having positive energy densities are ensured for the range $\nu<1$. 

\begin{table*}[htp]
\centering  
\caption{critical points of the autonomous system described by Eq.~(\ref{sec3:eqn3}) for the bulk viscosity model $s=1/2$ along with the conditions of existence and acceleration expansion. The effective equation of state parameter has been also included.}
\begin{ruledtabular}
\begin{tabular}{ccccccccccc}
Point & $\Omega_{r}$ & $\Omega_{m}$ & $\Omega_{\rm vac}$ & $w_{\rm eff}$ & \text{Existence} & \text{Acceleration}
  \\ \hline
 $(\rm Ia)$&$1-\nu$&$0$ &$\nu$ &$\frac{1}{3}(1-4\nu)$ & $\forall \nu,\hat{\xi}_{0}$ & $\text{No}$\\
 $(\rm Ib)$&$0$
 &$1-\nu$& $\nu$ & $-\nu \pm \sqrt{1 - \nu} \hat{\xi}_{0}$ & $ \nu<1, \forall\hat{\xi}_{0}$ & $\text{Yes}\; (\text{see main text})$\\
 $(\rm Ic)$&$0$
 &$\hat{\xi}_{0}^{2}$& $1-\hat{\xi}_{0}^{2}$ & $-1$ & $\forall \nu,\hat{\xi}_{0}$ & $\text{Yes}$\\
 $(\rm Id)$&$1-\nu-9\hat{\xi}_{0}^{2}$
 &$9\hat{\xi}_{0}^{2}$& $\nu$ & $\frac{1}{3}(1-4\nu)$ & $\forall \nu,\hat{\xi}_{0}$ & $\text{No}$\\
 $(\rm Ie)$&$0$
 &$0$& $1$ & $-1$ & $\forall \nu,\hat{\xi}_{0}$ & $\text{Yes}$\\
\end{tabular}
\end{ruledtabular}\label{M1:critical points}
\end{table*}

On the other hand, model parameters can play a crucial role in determining the stability conditions of the model. This can be checked by setting the right sign of the real parts of the eigenvalues associated to the Jacobian matrix of the the linear system. So the dynamical character of the critical points are displayed in table \ref{M1:eigenvalues}. If the conditions $\nu,\hat{\xi}_{0}\ll 1$ are taken beforehand, the stability criteria do not change considerably compared to the $\Lambda$CDM model. Let us be however more flexible to be able to infer the whole range of the model parameters from stability arguments. This is reported in table \ref{M1:eigenvalues} where we have introduced for short notation the quantities $\beta=1-\hat{\xi}_{0}^2$ and $\chi=1-9\hat{\xi}_{0}^2$ in the stability conditions.  It is worthwhile noting that the sign of $\hat{\xi}_{0}$ is crucial for determining the dynamical character of the critical point $(\rm Ie)$, leaving no room for the parameter space that satisfies the requirement $\lambda_{1},\lambda_{2}<0$ to be an attractor for $\hat{\xi}_{0}>0$. This means that even thought the late-time acceleration can be driven by this (unstable) point, the universe will depart from this stage due to bulk viscosity effects. At some point the trajectory will reach the true stable de-Sitter solution $(\rm Ic)$ where the bulk viscosity is also present, or the phantom-like solution $(\rm Ib)$ if the strong viscous regime is considered instead. So it is possible that the universe experiences two accelerated expansion stages or a single one driven by the critical points $(\rm Ib)$ or $(\rm Ic)$. There is no doubt from here that the effects of bulk viscosity on the background cosmological dynamics are meaningful: on one side it spoils the stability of $(\rm Ie)$, but on the other side it shapes suitable conditions for ensuring stable solutions. This is indeed the sharpest distinction between this model and the $\Lambda$CDM model at the background level.
\begin{table*}[htp]
\centering  
\caption{Eigenvalues and stability conditions that set the dynamical character of the associated critical points for model 1.}
%
\begin{ruledtabular}
\scalebox{0.95}{
\begin{tabular}{cccl}
\multicolumn{1}{c}{Point} & \multicolumn{1}{c}{$\lambda_{1}$} & $\lambda_{2}$ & \multicolumn{1}{c}{\text{Stability}} \\ \hline
$(\rm Ia)$& $4$ &$ \infty$ &\multicolumn{1}{c}{$\text{Repeller}\; \forall \nu, \hat{\xi}_{0}>0$}\\
 $(\rm Ib)$&$-1 + \nu \mp 3 \sqrt{1 - \nu} \hat{\xi}_{0}$&$-3 (-1 + \nu \pm \sqrt{1 - \nu} \hat{\xi}_{0})$ &$(-):\text{saddle} \;\text{if}\; \nu<\beta; \text{attractor} \;\text{if}\; \beta<\nu<1\;\forall\;\hat{\xi}_{0}>0$\\
& & &$(+):\text{saddle} \;\text{if}\; \nu<\chi; \text{repeller} \;\text{if}\; \beta(\chi)<\nu<1\;\forall\;\hat{\xi}_{0}>0$\\
 $(\rm Ic)$&$4 (-1 + \nu)$&$\frac{3}{2} (-1 + \nu + \hat{\xi}_{0}^{2})$ &$ \text{Repeller} \;\text{if}\;  \nu>1; \text{saddle} \;\text{if}\; \beta<\nu<1; \text{attractor} \;\text{if}\; \nu<\beta\; \forall\;\hat{\xi}_{0}>0$\\
 $(\rm Id)$&$4 - 4 \nu$&$-\frac{1}{2} (-1 + \nu + 9 \hat{\xi}_{0}^{2})$ &$\text{Repeller} \;\text{if}\;  \nu<\chi; \text{saddle} \;\text{if}\; \chi<\nu<1; \text{attractor} \;\text{if}\; \nu>1\; \forall\;\hat{\xi}_{0}>0$\\
 $(\rm Ie)$&$-4$ &$\infty$ &\multicolumn{1}{c}{$\text{Saddle}\; \forall \nu, \hat{\xi}_{0}>0$}\\
\end{tabular}
}
\end{ruledtabular}\label{M1:eigenvalues}
\end{table*}

\begin{figure*}
\centering
\includegraphics[width=0.47\hsize,clip]{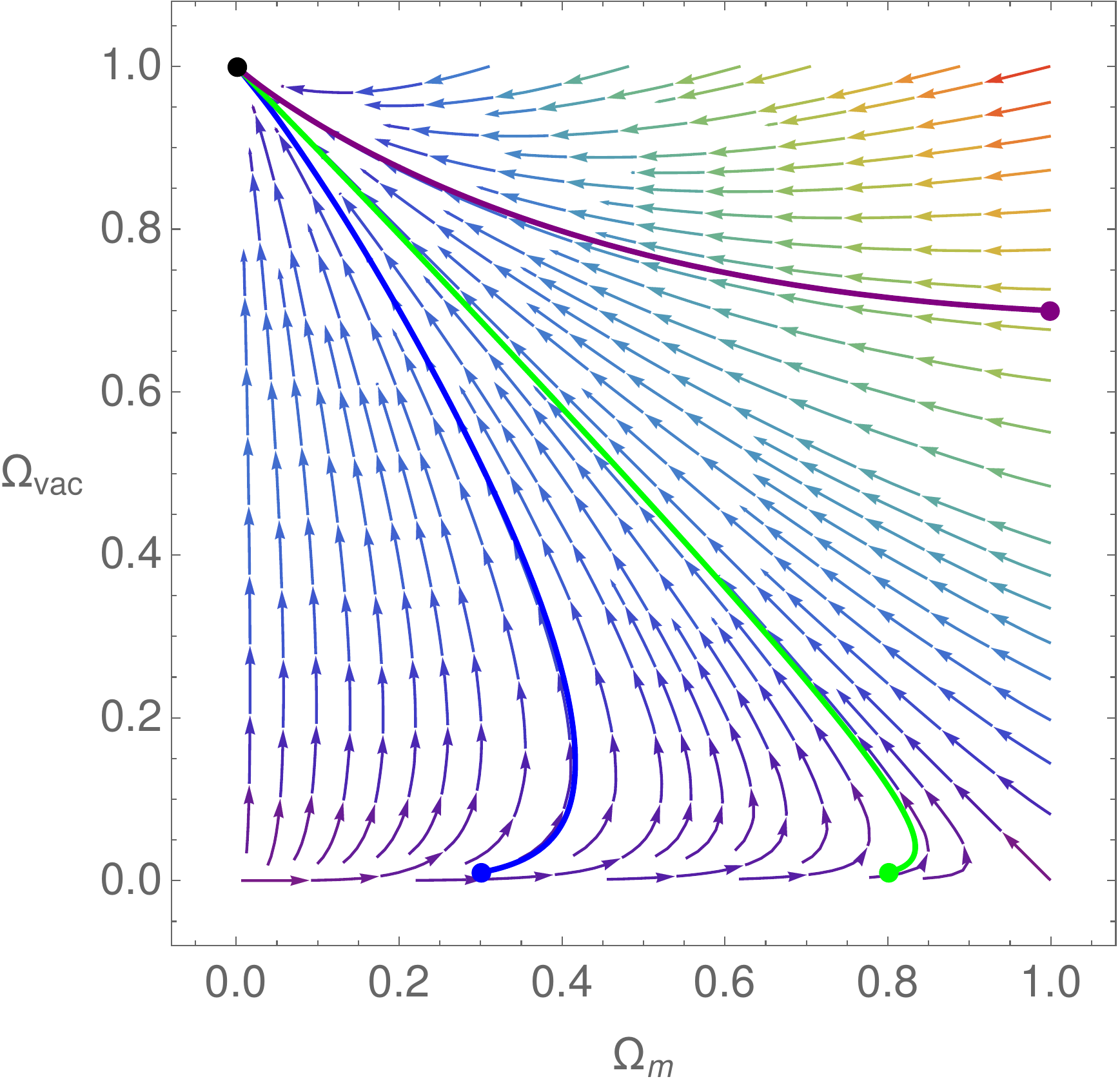}
\ \ \
\includegraphics[width=0.47\hsize,clip]{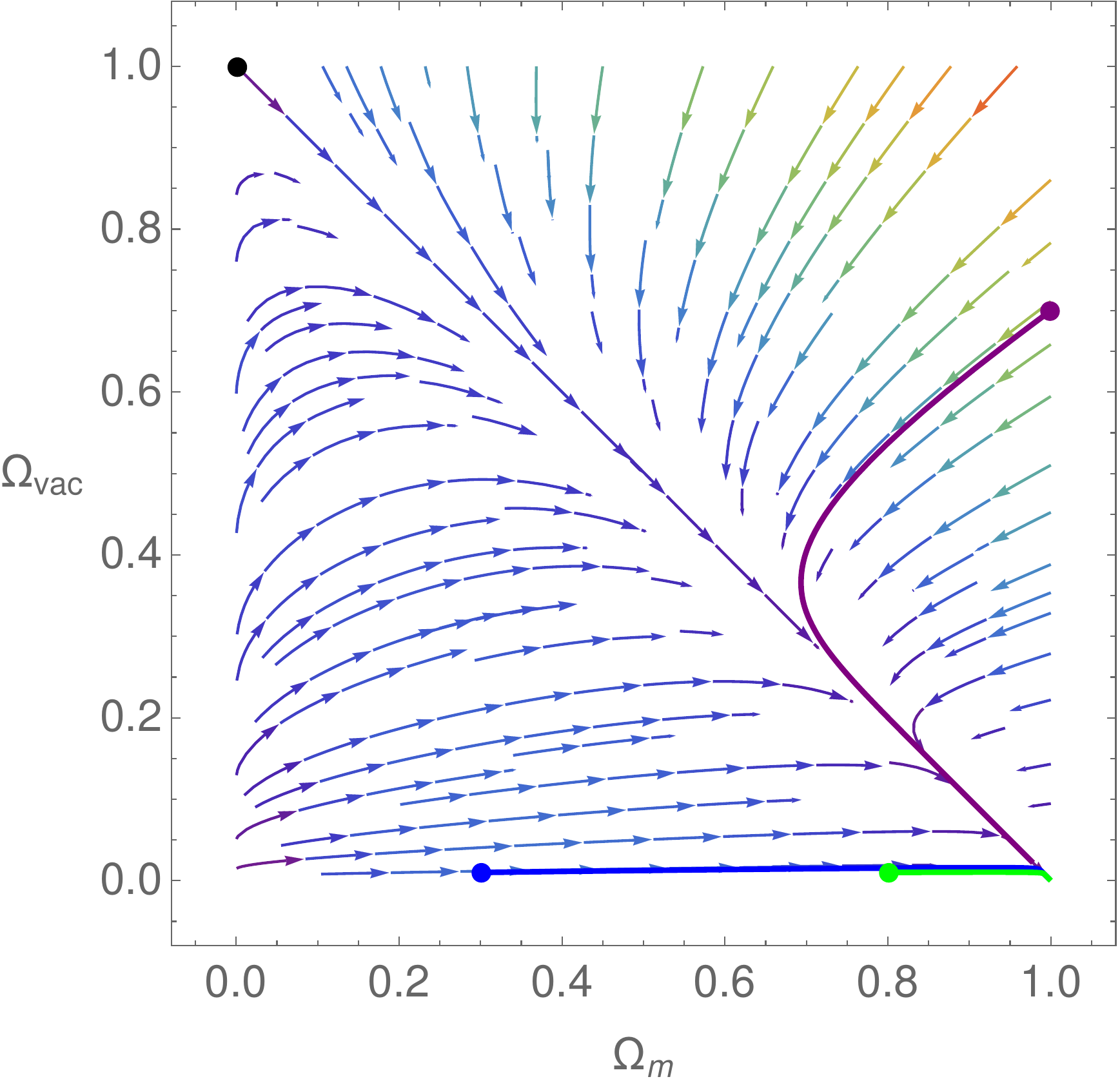}
\caption{Phase space of the system along with some trajectories for different initial conditions: 
($\Omega_{m}^{(i)}= 0.3,\; \Omega_{\rm vac}^{(i)}= 0.01$); ($\Omega_{m}^{(i)}= 0.8,\; \Omega_{\rm vac}^{(i)}= 0.01$);
($\Omega_{m}^{(i)}= 1,\; \Omega_{\rm vac}^{(i)}= 0.7$), for blue, green and purple curves, respectively.  Left panel evidences the attractor character of the system after evolving with $\nu=\hat{\xi}_{0}=10^{-4}$ while right panel shows the saddle-like behavior for $\nu=10^{-4}$ and $\hat{\xi}_{0}=1.05$.} \label{fig:M1}
\end{figure*}

Some numerical trajectories are also displayed in the two-dimensional phase space $(\Omega_{m},\Omega_{\rm vac})$ in fig.~\ref{fig:M1} for different initial conditions as explained in the caption, they all leading to the attractor point $(0,1)$ $(\rm Ic)$, i.e. to an universe experiencing an accelerated expansion after passing close to a saddle point describing dark matter domination (see left panel). This case corresponds to the case $\hat {\xi}_{0}=10^{-4}$ while right panel shows the strong viscous case $\hat{\xi}_{0}\sim \mathcal{O}(1)$ that illustrates the fact that $\hat{\xi}_{0}>1$, keeping $\nu \ll 1$, changes the dynamics of the critical point $(\rm Ib)$ from saddle to attractor point. Whether such large values correspond to a realistic viscosity scenario of dark matter, without invoking the unified dark sector description, is a theme that must be independently assessed from cosmological parameter estimation when calculating the best-fit parameters from observational data. This is also a subject that must be treated in the future.

Other values of $s$ for this first class of models are briefly discussed here as well as their main features. For instance, the case $s=0$, which leads to the well known parameterization $\xi\sim H$, can not provide a radiation domination period: there do not exist any physical conditions such that $\Omega_{r}\neq0$ along the entire phase space trajectories. Hence, this case must be discarded as a suitable cosmological solution. Before going forward let us, however, describe two critical points that are also present for others values of $s$.
The first point corresponds to $\Omega_{m}=1-\nu$ and $\Omega_{\rm vac}=\nu$ with effective equation of state $w_{\rm eff}=-\nu- \hat{\xi}_{0}$, which describes matter domination provided that $\nu\ll 1$, which can generate an accelerated expansion if $\nu\leq1/3$ and $\hat{\xi}_{0}>\frac{1}{3}(1-3\nu)$ (strong viscous regime). This critical point is analogue to the critical point ($\rm Ib$) but with a different equation of state. The second critical point we find is nothing more than a duplicate critical point as described by the point ($\rm Ic$). 

The case $s=1$ is also phenomenologically interesting because the viscous fluid features are present in one of the critical points with the magnitude of $\hat{\xi}_{0}$ (and the sign of $\nu$) determining unequivocally the cosmological behavior of this point. That is to say, $w_{\rm eff}=\nu (-1 +\hat{\xi}_{0}) - \hat{\xi}_{0}$. So accelerated expansion is possible provided that  $\hat{\xi}_{0}>1$ and $\nu < \frac{-1+3\hat{\xi}_{0}}{-3+3\hat{\xi}_{0}}$ yielding thus a phantom-like behavior, or simply $\hat{\xi}_{0}=1$ leading to a de de-Sitter-like solution. The first conditions implies clearly that $\nu>0$ and $\hat{\xi}_{0}>0$. For such values of $\hat{\xi}_{0}$ the model is in the strong viscous regime which allows the dark matter component to drive the current expansion of the universe through the bulk viscosity effect. Notice that the parameter allows the vacuum energy component to exist during this period despite that it does not play any role in the acceleration: $\Omega_{\rm vac}=\nu$ ($\Omega_{m}=1-\nu$). This point is hence an attractor when $\nu<1$ and $\hat{\xi}_{0}>1$, and saddle for the same range of $\nu$ and $0<\hat{\xi}_{0}<1$. The latter condition breaks clearly the strong viscous regime, necessary to realize accelerated expansion, whereby this point corresponds, for such parameter values,  to standard dark matter domination in this weak regime. We can conclude from phase space analysis that this point is quite appealing to cosmological dynamics of the late universe, and, as demanded, it must be put under scrutiny with the help of observational data to ensure its cosmological viability. 

Another commonly unexplored choice is $s=-1$, but it is ruled out, as for the $s=0$-case, because the critical point that describes the radiation era is not real valued. Larger positive values of $s$ are also cosmologically viable with one of their critical points characterized by a common effective equation of state and written in the general way as $w_{\rm eff}=-\nu\pm(1-\nu)^{s}\hat{\xi}_{0}$ where the branch $-$ corresponds to even integers only: $(-1)^{s+1}$; and the branch $+$ for all the others, including half-integers. Further exploration about suitable power law values is beyond the scope of this paper since there is not (as far as we know) a guidance criterion from physical grounds as thermodynamics principles, apart from dynamical system analysis, to select particular viscous models. The reader can find a more detail discussion about the general pattern of the critical points due to this new parameterization of the bulk viscosity in reference \cite{Gomez:2022qcu}.

It is worthwhile mentioning that the critical points (Ia) and (Ie) of Table I do not correspond to perturbative fixed points, and therefore the linear stability analysis performed by computing their eigenvalues and associated properties cannot be trusted. This technical involved issue will usually not be mentioned as the validity of the standard stability analysis beyond the linear contributions relies on Malkin's nonlinear stability theorem \cite{Malkin_52}. Nevertheless, both points (Ia) and (Ie) lead to stationary points of the dynamical system describing the model and therefore they can be included as critical points, but taking care of their stability properties by numerical analysis.    

\subsection{\label{sec:Model2Dynamical} Model 2: $\tilde{\nu}=\nu/2$ and $s=1$}
This sub-class of model with $\tilde{\nu}=\nu/2$ is characterized by $s=1$ in Eq.~(\ref{sec3:eqn4}). The set of 4 critical points associated to this system is reported in Table \ref{M2:critical points}. There exists one trajectory in phase space describing the background cosmological dynamics that can follow the standard radiation dominated ($\rm{IIa}$) and the accelerated expansion ($\rm{IIb}$) stages similar to the $\Lambda$CDM model.  The intermediate period is described however by non-standard dark matter ($\rm{IIc}$) which involves both the effects of the running and the viscous dark matter fluid feature. This happens particularly for values where $\nu, \hat{\xi}_{0}\ll 1$. Turning off the piece of the running vacuum energy density associated to $\nu$ the viscosity effect remains hidden in the energy densities but encoded in the effective equation of state ( $w_{\text{eff}}=-\hat{\xi}_{0})$ and in the stability conditions as can be inferred in Tables \ref{M2:critical points} and \ref{M2:eigenvalues}, respectively.
One may in principle argue that if $\hat{\xi}_{0}$ is small enough, matter domination, as we expect, may be realized. Turning on the $\nu$-parameter does not provide a successful exit either to this problem. This point will be investigated numerically using a high-precision solver during the numerical evolution. 

In the most general case this point can describe accelerated expansion ($w_{\rm eff}<-1/3$) provided that the conditions $0 < \hat{\xi}_{0} \leq 1/3$ and $\frac{-2 + 6 \hat{\xi}_{0}}{-3 + 3 \hat{\xi}_{0}} < \nu < 1$ are fulfilled. Here we have taken in advance the constraint $0<\hat{\xi}_{0}<1$. 
In the strong viscous regime, this point can also generate accelerated expansion  similar to the point ($\rm{Ib}$) of model 1 either in the absence of running ($\nu=0$) or in the most general case $\nu\neq0$ and $\hat{\xi}_{0}\geq1/3$. So in the strong viscous regime, the dynamical character of this point is once more changed from saddle to attractor. In both cases the bulk viscosity determines the dynamical character of the expansion. In the  case $\nu=0$ for instance, one finds simply $w_{\rm eff}=-\hat{\xi}_{0}$. So depending of the bulk viscosity strength with $\hat{\xi}_{0}>0$, the acceleration can reveal different behaviors including the well-know phantom-like and  de-Sitter ($\hat{\xi}_{0}=1$) solutions. This is also true in the general case as along as the bulk viscosity is the dominant effect. To see this, let us take the suggestive value $\hat{\xi}_{0}=1$. This yields $\Omega_{m}=1-\nu$, $\Omega_{\rm vac}=\nu$ and $w_{\rm eff}=-1$. 

From construction one expects however $\nu\ll 1$ ($\Omega_{m}\to 1$) such that the strong bulk viscosity regime, is once again capable of pushing away the accelerated expansion of the universe instead of conventional mechanisms of dark energy. These results are nothing more that bulk viscous unified scenarios of dark matter and dark energy. So, neglecting completely the running effect, viscous dark matter can help to  describe independently the complete cosmological dynamics
under the underlying physical mechanics behind bulk viscosity. Notice that no matter the dynamical character of this critical point their associated energy density parameters must be positive define which leads to the weak constraint  $0\leq\nu\leq1$ and $\hat{\xi}_{0}>0$. 

On the other hand, we report the last critical point ($\rm{IId}$) that can be surprisingly standard radiation domination for $\nu = \frac{1 + 3 \hat{\xi}_{0}}{3 \hat{\xi}_{0}}$. This point can also generate accelerated expansion by combining both the running and bulk viscosity effects. For instance, the de-Sitter solution is realizable here taking $\hat{\xi}_{0}=1$ necessarily. The condition for accelerated expansion is achieved even in the weak viscosity regime $\hat{\xi}_{0}\ll1$ as along as the general condition  $\hat{\xi}_{0}>\frac{3\nu}{-4+6\nu}$ is fulfilled. The sign of $\nu$ is decisive to set the dynamical character of the
expansion. For instance, for a given positive $\nu$ and derived $\hat{\xi}_{0}$, the expansion is a phantom-like while negative $\nu$ leads to $w_{\rm eff}>-1$. On the other hand, the condition of positive energy densities put the very tight constraint $\frac{1+3\nu}{-3+6\nu}<\hat{\xi}_{0}<1$ where $\nu>4/3$, which is compatible with the less restrictive condition for acceleration expansion but far beyond the expected value from physical grounds. So this solution can not describe successfully the current accelerated expansion whereby it is not of physical interest in this form. Notice finally that it is not possible to neglect the running vacuum energy density or the bulk viscosity here due to the conditions of existence for this critical point that prevent both $\nu$ and $\hat{\xi}_{0}$ from nullity.

In the case of vanishing bulk viscosity $\hat{\xi}_{0}=0$, there only appear the first three critical points where ($\rm{IIc}$) is reduced to $\Omega_{m}=\frac{4 (-1 + \nu)}{-4 + 3 \nu}$ and $\Omega_{\rm vac}=\frac{\nu}{4 - 3 \nu}$. Demanding positive energy density parameters yields the constraint $0<\nu<1$ and the existence of the critical point itself imposes  $\nu\neq\frac{4}{3}$. Notice that negative values are not allowed from this simple request which is in consistent with the observational limit inferred by cosmological data. We remind that this point accounts for matter domination and exhibits small deviation from the $\Lambda$CDM model due to the presence of the  running vacuum energy density. This is indeed the only difference at the background level that can be appreciated from phase space analysis. It is interesting the early presence of the vacuum energy density in this period, like in the most general form of this class of models ($\hat{\xi}_{0}\neq0$), which is appealing into the light of the coincidence problem and, presumably, into the mechanism behind the formation of large scale structure in the universe. This latter aspect must be examined carefully to find more compelling distinctions beyond the cosmological background.  

\begin{table*}[htp]
\centering  
\caption{critical points of the autonomous system described by Eq.~(\ref{sec3:eqn4}) for the bulk viscosity model $s=1$ along with the conditions of existence. The effective equation of state parameter has been also included.}
\scalebox{0.82}{
\begin{tabular}{ccccccccccc}
\hline
\hline
Point & $\Omega_{r}$ & $\Omega_{m}$ & $\Omega_{\rm vac}$ & $w_{\rm eff}$ & \text{Existence} & \text{Acceleration}
  \\ 
  \hline
 $(\rm IIa)$&$1$&$0$ &$0$ &$\frac{1}{3}$ & $\forall \nu,\hat{\xi}_{0}$ & $\text{No}$\\
 $(\rm IIb)$&$0$
 &$0$& $1$ & $-1$ &$ \forall \nu,\hat{\xi}_{0}$ & $\text{Yes}$\\
 $(\rm IIc)$&$0$
 &$\frac{4(1 - \nu)}{4 + 3 \nu (-1 + \hat{\xi}_{0})}$& $\frac{\nu + 3 \nu \hat{\xi}_{0}}{4 + 3 \nu (-1 + \hat{\xi}_{0})}$ & $\frac{\nu (-1+\hat{\xi}_{0})-4 \hat{\xi}_{0}}{4 + 3 \nu (-1 + \hat{\xi}_{0})}$ & $\forall \nu,\hat{\xi}_{0}\neq 1-\frac{4}{3\nu}$ & $\text{Yes}\; $
 \\
  $(\rm IId)$&$\frac{(-1 + \hat{\xi}_{0}) (1 + 
   3 (1 - \nu + \nu^{2})\hat{\xi}_{0})} {\nu \hat{\xi}_{0}(1 - 3 \nu + 3 \hat{\xi}_{0}) }$
 &$\frac{-4 - 12\hat{\xi}_{0} + 12 \nu \hat{\xi}_{0}}{3 \nu \hat{\xi}_{0}(-1 + 3 \nu - 3 \hat{\xi}_{0})}$& $\frac{(1 - 3 (-1 + \nu) \hat{\xi}_{0})^{2}} {3 \nu\hat{\xi}_{0} (-1 + 3 \nu - 3 \hat{\xi}_{0})}$ & $-1-\frac{4 \nu (-1 +  \hat{\xi}_{0})}{-1+3 \nu - 3 \hat{\xi}_{0}}$ & $\nu\neq0\neq\hat{\xi}_{0}, \hat{\xi}_{0}\neq (-\frac{1}{3}+\nu)$& $\text{Yes}\; $
 \\
 \\
 \hline
 \hline
\end{tabular}\label{M2:critical points}
}
\end{table*}

As to the stability conditions for this model, compatible with $\hat{\xi}_{0}>0$ (see Table \ref{M2:eigenvalues}), they are plainly achieved. The critical point ($\rm{IIa}$) may in principle be repeller or saddle in the general situation. Nevertheless, imposing the criteria $\hat{\xi}_{0}>0$ and $\nu$ small, this critical point must be necessarily a repeller. The resulting de-Sitter solution for this model ($\rm{IIb}$) is stable for the large range $-1<\nu<1$ and $0<\hat{\xi}_{0}<1$. Physical expectations however tell us that $\nu\ll1$. The critical point ($\rm{IIc}$), in the form describing matter domination ($\nu,\hat{\xi}_{0}\ll1$), is a saddle point because their associated eigenvalues have always opposite signs by the requirement $\hat{\xi}_{0}>0$ within the same allowed range of the parameter space as critical point ($\rm{IIb}$). This same critical point can be also attractor as discussed and its associated (reals parts of) eigenvalues are both negatives  for the range $-1<\nu<0$ and $\hat{\xi}_{0}>0$. The sign of $\nu$ is crucial for ensuring the stability. Lastly, the critical point ($\rm{IId}$) is sensitive to the sign of both $\nu$ and $\hat{\xi}_{0}$ whereby we have chosen the appropriated sign ($\nu>0$), by numerical examination, so that the point is an attractor. Notice that we have used the abbreviated quantity
\begin{widetext}
\begin{eqnarray}
    \chi&\equiv \nu (4 + 3 \nu (-1 + \hat{\xi}_{0})) (-1 + \hat{\xi}_{0})\hat{\xi}_{0} (4 + 36 \hat{\xi}_{0} + 3 (-4 \nu + (-2 + \nu) \nu (16 - 6 \nu + \\ \nonumber &9 \nu^2) \hat{\xi}_{0}+  6 (6 + \nu (-14 + 3 \nu (6 + (-4 + \nu) \nu))) \hat{\xi}_{0}^2 + 9 (2 + (-2 + \nu) \nu)^2 \hat{\xi}_{0}^3)),\label{sec3:eqn6}
\end{eqnarray}
\end{widetext}
in the the eigenvalues. For $\nu>0$ the effective equation of state has a phantom-like behavior according to the requirement $\nu,\hat{\xi}_{0}\ll 1$.

\begin{table*}[htp]
\centering  
\caption{Eigenvalues and stability conditions for determining the dynamical character of the associated critical points for model 2.}
\scalebox{0.82}{
\begin{tabular}{ccccccccccc}
\hline
\hline
Point & $\lambda_{1}$ & $\lambda_{2}$ & \text{Stability} \\ \hline
$(\rm IIa)$&$4$&$\frac{4 - 12 (-1 + \nu)\hat{\xi}_{0}}{4 + 3 \nu (-1 + \hat{\xi}_{0})}$ &$\text{Repeller} \; \text{if}\; 0 \leq \nu \leq 1 \land \ \hat{\xi}_{0} > 0$\\
 $(\rm IIb)$&$-\frac{12 (-1 +\nu) (-1 +\hat{\xi}_{0})}{4 + 3 \nu (-1 + \hat{\xi}_{0})}$&$-4$ &$\text{Attractor} \;\text{if}\; -1 < \nu < 1 \land 0 < \hat{\xi}_{0}< 1$\\
 $(\rm IIc)$&$\frac{12 (-1 +\nu) (-1 +\hat{\xi}_{0})}{4 + 3 \nu (-1 + \hat{\xi}_{0})}$ &$-\frac{4 (1 + 3 (1 + (-1 + \nu) \nu) \hat{\xi}_{0})}{ 4 + 3 \nu (-1 +\hat{\xi}_{0})}$ &$\text{Saddle}: 0 \leq \nu < 1 \land 0 < \hat{\xi}_{0}< 1;\; \text{attractor}: -1 < \nu < 0 \land \hat{\xi}_{0}>0$\\
 $(\rm IId)$&$\frac{-6 \nu^2 (4 + 3 \nu (-1 +\hat{\xi}_{0})) (-1 + \hat{\xi}_{0}) \hat{\xi}_{0} - 
   2 \chi}{\nu (4 + 3\nu (-1 + \hat{\xi}_{0})) (-1 + 3 \nu - 3 \hat{\xi}_{0})\hat{\xi}_{0}}$&$\frac{2 (-3 \nu^2 (4 + 
       3\nu (-1 + \hat{\xi}_{0})) (-1 + \hat{\xi}_{0}) \hat{\xi}_{0} + \chi}{\nu (4 + 3\nu (-1 + \hat{\xi}_{0})) (-1 + 3 \nu - 3 \hat{\xi}_{0})\hat{\xi}_{0}}$ &$\text{Attractor} \;\text{if}\; 0<\nu<1/3\land 0<\hat{\xi}_{0}< 1$ \\\
       \\
\hline
\hline
\end{tabular}\label{M2:eigenvalues}
}
\end{table*}

In most of the critical points the requirement $\hat{\xi}_{0}>0$ (and reasonably small values to be still relevant) selects the specific region $\nu>0$ of the parameter space. Though this region of the parameter space is consistent with the demand for positive energy densities ($\nu>4/3$) we remind that this solution is physically attractive due to the  inferred large value of $\nu$. We conclude that phase space analysis along with the condition of positive energy densities do not allow accelerated solutions, apart from the de-Sitter solution ($\rm{IIb}$) and the one driven by bulk viscosity $\rm{IIc}$, where the running vacuum energy density is the main agent responsible for the expansion.
In the simplest version of this class of running vacuum models, that is $\hat{\xi}_{0}=0$, stability of the resulting solutions are plainly achieved for the range $\nu<1$ which is consistent with the one demanded for having positive energy densities. The phase space analysis therefore left a few suitable critical points to describe the cosmological backgrounds dynamics. These critical points are slightly different to the ones found in model 1 either in the limit $\nu,\hat{\xi}_{0}\ll1$ or in the strong viscous regime $\hat{\xi}_{0}>1$. For this reason their respective phase spaces are practically indistinguishable from each other. So numerical plots are not shown here. Notice however that the parameter space are distinct, in particular model 1 does allow $\nu<0$.

\subsection{\label{sec:Model3Dynamical} Model 3: $\tilde{\nu}=\nu/2$ and $s=1/2$}\label{systemmodel3}
This model corresponds to $s=1/2$ for the bulk viscosity exponent in the system Eq.~(\ref{sec3:eqn4}). Some common solutions to the already discussed ones are found, like solutions ($\rm IIa$) and ($\rm IIb$) of model 2 and ($\rm Ic$) of model 1, as well as the sub-manifolds belonging to the branches $\nu=0$ and $\hat{\xi}_{0}=0$, so we report the two different solutions in Table \ref{M3:critical points} for a each given sign, and discuss their main physical properties as follows. The most interesting feature of taking the Ansatz Eq.~(\ref{sec2:eqn5}) is that this allows the existence of a (viscosity-running) two-parameters family of  solutions (IIIa) whose respective EoS coincide with the one of point ($\rm IIc$) in the $\hat{\xi}_0 \to 0$ limit, despite they were deduced for different bulk viscosity exponents.

Specifically, the solution ($\rm IIIa$) represents a general form of matter domination solution in the sense that this covers the limit cases: $w_{\rm eff}=\pm\hat{\xi}_{0}$ when $\nu\to0$ and $w_{\rm eff}=\frac{\nu}{4-3\nu}$ when $\hat{\xi}_{0}\to0$. This point can also describe accelerated expansion whose effective equation of state takes a more involved form (see Table \ref{M3:critical points}). The necessary condition for accelerated expansion can be very well approximated to $\hat{\xi}_{0}\gtrsim1/3$ and $0<\nu<1$ for the negative branch. For the positive one however there is not a suitable range of the parameter space, fulfilling particularly $\nu\ll1$, that provides the cosmic acceleration. We have defined the parameter $\eta=64 - 112 \nu + \nu^2 (48 + 9 \hat{\xi}_{0}^2)$ everywhere for the sake of compactness. Other general solutions are ruled out by demanding positive energy density parameters or because they do not respect the physical condition $\nu\ll1$.

\begin{table*}[htp]
\centering  
\caption{critical points of the autonomous system described by Eq.~(\ref{sec3:eqn4}) for the bulk viscosity model $s=1/2$ along with the conditions of existence. The effective equation of state parameter has been also included.}
\begin{ruledtabular}
\begin{tabular}{ccccccccccc}
Point & $\Omega_{r}$ & $\Omega_{m}$ & $\Omega_{\rm vac}$ & $w_{\rm eff}$ & \text{Existence} & \text{Acceleration}
  \\ \hline
 $(\rm IIIa)$&$0$&$\frac{(3 \nu \hat{\xi} \pm \eta^{1/2})^{2}}{(-8 + 6 \nu)^2}$ &$\frac{\nu (8 - 3 \nu (2 + 3 \hat{\xi}_{0}) \mp 
   3 \hat{\xi}_{0} \eta^{1/2})}{2 (4 - 3 \nu)^2}$ &$\frac{3 \nu^2 + \nu (-4 + 6 \hat{\xi}_{0}^{2}) \pm 2 \hat{\xi}_{0} \eta^{1/2}}{(4 - 3 \nu)^2}$ & $ \nu\neq\frac{4}{3},\nu\leq1,\hat{\xi}_{0}>0$ & $\text{Yes}$\\
\end{tabular}
\end{ruledtabular}\label{M3:critical points}
\end{table*}

It is interesting to note that the solution ($\rm Ic$) of model 1 is also a solution of the present model. Eigenvalues however change naturally the form due to the structure of the system but there is an ample region of the parameter space, $\nu<1$ and $-\sqrt{1 - \nu} < \hat{\xi}_{0} < \sqrt{1 - \nu}$, for which accelerated expansion can be still driven by the bulk viscosity effect in this general scenario. Out of this region, the point corresponds to standard dark matter domination. On the other hand, the de-Sitter solution ($\rm IIb$) is plainly preserved for $\nu\neq0$ within this case. 

Eigenvalues for the renewed solutions are too lengthy to be reported, so we have to check numerically this aspect to establish the dynamical character of those solutions. For the solution ($\rm IIa$) with negative branch, the resulting parameter space reads approximately $\hat{\xi}_{0}>1$ (strong viscous regime) and  $|\nu|<1$. This is strictly valid for $\nu>\mathcal{O}(\pm10^{-2})$. The positive branch corresponds to a repeller for the physical parameter space $0<\hat{\xi}_{0}<1$ and $0<\nu<1$.

\subsection{\label{sec:Model4Dynamical} Model 4: $\tilde{\nu}=\nu/2$ and $s=0$}
This is also a sub-class of model belonging to $\tilde{\nu}=\nu/2$ but with $s=0$ in Eq.~(\ref{sec3:eqn4}), which leads to functional dependence $\hat{\xi}\sim H$ for the bulk viscosity coefficient. We have found in a previous work without including the running effects that this exponent is discarded because it can not describe consistently the whole cosmological evolution of the universe: radiation dominated period is absent as a critical point. Can the running vacuum energy density effects restore the goodness that offer, for instance, the $\Lambda$CDM in this regard? unfortunately the answer is not. So, this model is not of cosmological interest. For completeness we report however the set of new critical points associated to this system in Table \ref{M7:critical points}. The first critical point listed $(\rm IVa)$ is similar to the point ($\rm{IC}$) of model I (with $s=1/2$), providing also de-Sitter-like acceleration expansion with non-vanishing bulk viscosity. The point $(\rm IVb)$ corresponds to matter domination with non-vanishing dark energy density thanks to both $\nu$ and $\hat{\xi}$. This is a kind of scaling solution. When turning off the running effects, it was shown that negative integers of the exponent $s$ can not provide, by any means, radiation domination. Nevertheless, for the present model with a (general) negative real $s$-value, it leads to a higher non-lineal differential equation that are difficult to solve by the methods employed in this work. Though we expect that such harmful features are propagated by the fact of taking $\hat{\xi}\sim H$, we can not discard certainly this possibility.

\begin{table*}[htp]
\centering  
\caption{critical points of the autonomous system described by Eq.~(\ref{sec3:eqn4}) for the bulk viscosity model $s=0$ along with the conditions of existence. The effective equation of state parameter has been also included.}
\begin{ruledtabular}
\begin{tabular}{ccccccccccc}
Point & $\Omega_{r}$ & $\Omega_{m}$ & $\Omega_{\rm vac}$ & $w_{\rm eff}$ & \text{Existence} & \text{Acceleration}
  \\ \hline
 $(\rm IVa)$&$0$&$\hat{\xi}$&$1-\hat{\xi}$&$-1$&$\forall \hat{\xi},\nu$&$\text{Yes}$\\
 $(\rm IVb)$&$0$&$\frac{-4 + \nu (4 + 3 \hat{\xi})}{-4 + 3 \nu}$&$\frac{\nu + 3 \nu \hat{\xi}}{4 - 3 \nu}$&$\frac{\nu + 4 \hat{\xi}}{-4 + 3 \nu}$&$\forall \hat{\xi}\;\land\;\nu\neq4/3$&$\text{No}$\\
\end{tabular}
\end{ruledtabular}\label{M7:critical points}
\end{table*}
%
\section{\label{sec:Numerical}Numerical Solutions}
In this section, we present the results obtained by the numerical integration of two sub-classes of running models studied in this paper, for specific elections of the exponent $s$ of the bulk viscosity coefficient (see Eq.~(\ref{sec2:eqn5})). Specifically, model 1 corresponds to the first class of models for which $\tilde{\nu}=0$ and $s=1/2$ are replaced into Eq.~(\ref{sec3:eqn3}), while models 2 and 3 correspond to  the second class model where $\tilde{\nu}=\nu/2$ and $s=1$ and $1/2$ respectively are inserted into Eq.~(\ref{sec3:eqn4}). Model 4, obtained from the second class model with $s=0$, is discarded due to the  physical argument explained in the subsection \ref{sec:Model4Dynamical}. 

For the numerical integration we have implemented an algorithm in the programming language \textit{Python}, using the \textit{solve\_ivp} module provided by the \textit{SciPy} open-source \textit{Python}-based ecosystem. The integration method chosen was \textit{RK45}, which is an explicit \textit{Runge-Kutta} method of order 5(4), with relative and absolute tolerances of $10^{-6}$ and $10^{-9}$, respectively. The systems of differential equations were integrated with respect to $N=\ln{a}$ (which is related to the redshift through the expression $1+z=a_{0}/a$), in the integration range of $-15\leq N\leq 5$, partitioned uniformly in $10\,000$ data points. 

It is worthwhile mentioning that, we integrate the differential equation for the Hubble parameter given by Eq.~(\ref{sec3:eqn5}) as well, moreover, we only integrate two of the dynamical equations for $\Omega_{r}$ and $\Omega_{m}$, as $\Omega_{\rm vac}$ can straightforwardly be obtained through the Friedmann's constraint of Eq.~(\ref{sec3:eqn2}). In this sense, the initial conditions of both class of models were chosen in order to match with the $\Lambda$CDM model at current time ($a_{0}=1$ or $z=0$) according to the Planck 2018 results \cite{Planck:2018vyg}, i.e., $H_{0}=100\frac{km/s}{Mpc}h$ where $h=0.674$, $\Omega_{m,0}=0.315$, and $\Omega_{r,0}=2.469\times 10^{-5}h^{-2}(1+0.2271N_{\rm eff})$ with $N_{\rm eff}=2.99$. Even more, all the numerical solutions where compared with their $\Lambda$CDM counterparts, considering that the expression for the $\Lambda$CDM Hubble parameter and effective equation of state parameter are given by
\begin{eqnarray}\label{LCDM}
    H(z)&=&H_{0}\sqrt{\Omega_{r,0}(1+z)^{4}+\Omega_{m,0}(1+z)^{3}+\Omega_{\Lambda,0}},\label{sec5:eqn1}\\
    \omega_{\rm eff}&=&\frac{4}{3}\Omega_{r}+\Omega_{m}-1,\label{sec5:eqn2}
\end{eqnarray}
where $\Omega_{\Lambda,0}=1-\Omega_{r,0}-\Omega_{m,0}$, $\Omega_{r}=\Omega_{r,0}(1+z)^{4}/E(z)^{2}$, and $\Omega_{m}=\Omega_{m,0}(1+z)^{3}/E(z)^{2}$, with $E(z)=H(z)/H_{0}$. 

All the figures shown in this section were obtained for two different types of combinations of the free parameters $\hat{\xi}_{0}$ and $\nu$. The first one consider combinations of $\hat{\xi}_{0}$ with only positive values of $\nu$, namely, $\hat{\xi}_{0}=1\times 10^{-4}$ and $\nu=5\times 10^{-4}$, $\hat{\xi}_{0}=9\times 10^{-3}$ and $\nu=5\times 10^{-4}$, $\hat{\xi}_{0}=9\times 10^{-3}$ and $\nu=1\times 10^{-2}$, and $\hat{\xi}_{0}=9\times 10^{-3}$ and $\nu=5\times 10^{-2}$. The second one consider combinations of $\hat{\xi}_{0}$ with only negative values of $\nu$, namely, $\hat{\xi}_{0}=1\times 10^{-4}$ and $\nu=-5\times 10^{-4}$, $\hat{\xi}_{0}=9\times 10^{-3}$ and $\nu=-5\times 10^{-4}$, $\hat{\xi}_{0}=9\times 10^{-3}$ and $\nu=-5\times 10^{-3}$, and $\hat{\xi}_{0}=9\times 10^{-3}$ and $\nu=-1\times 10^{-2}$. The figures are presented within the range $3.27\times 10^{6}<z+1\leq 0.1$, except for the model 3, for which the range $10^{5}\leq z+1\leq 0.1$ was used due to numerical difficulties.

\subsection{\label{NUmericalM1} Model 1}
\begin{figure*}
    \centering
    \subfigure[\label{fig:M1nu+Density}]{\includegraphics[scale = 0.35]{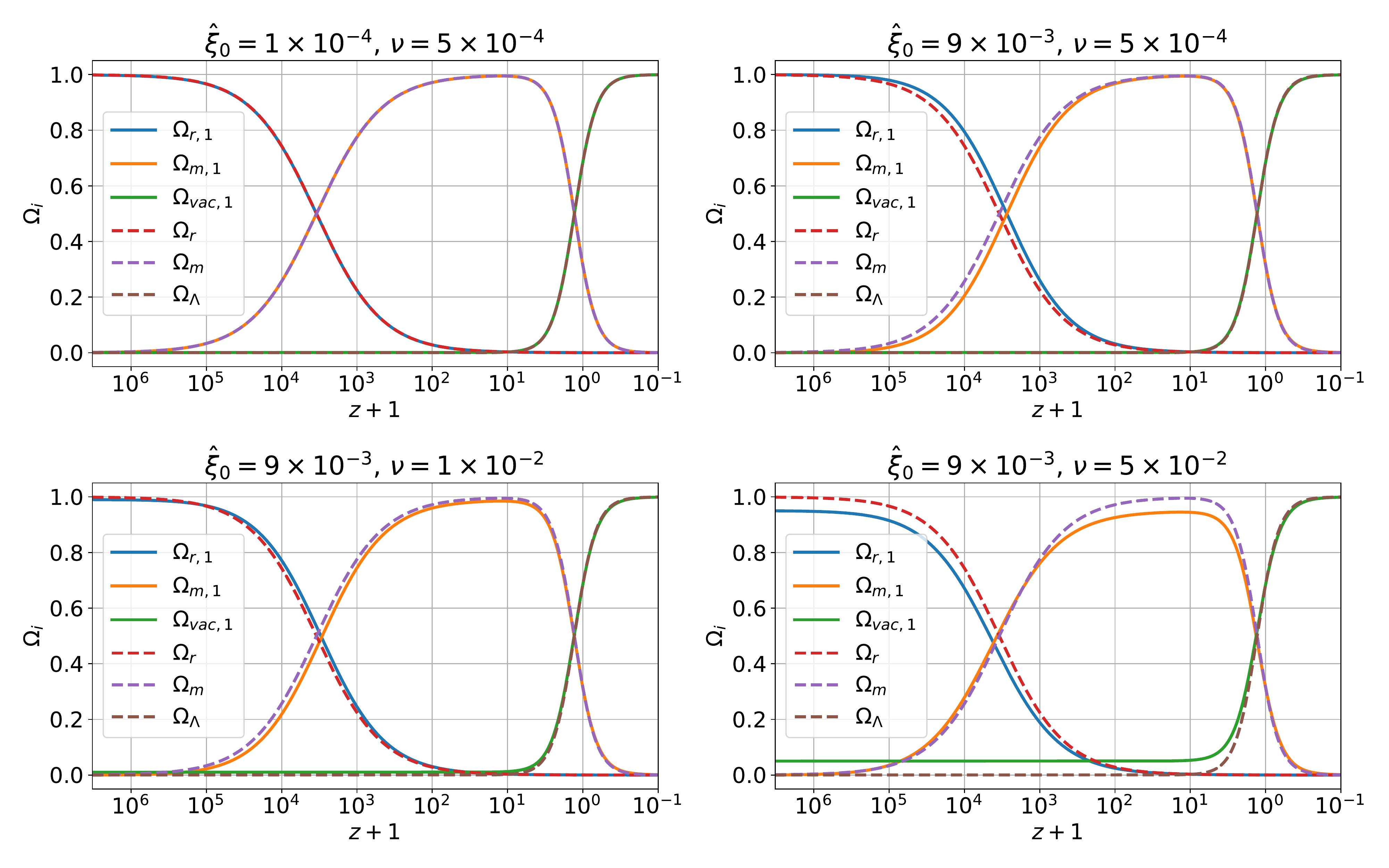}}
    \subfigure[\label{fig:M1nu-Density}]{\includegraphics[scale = 0.35]{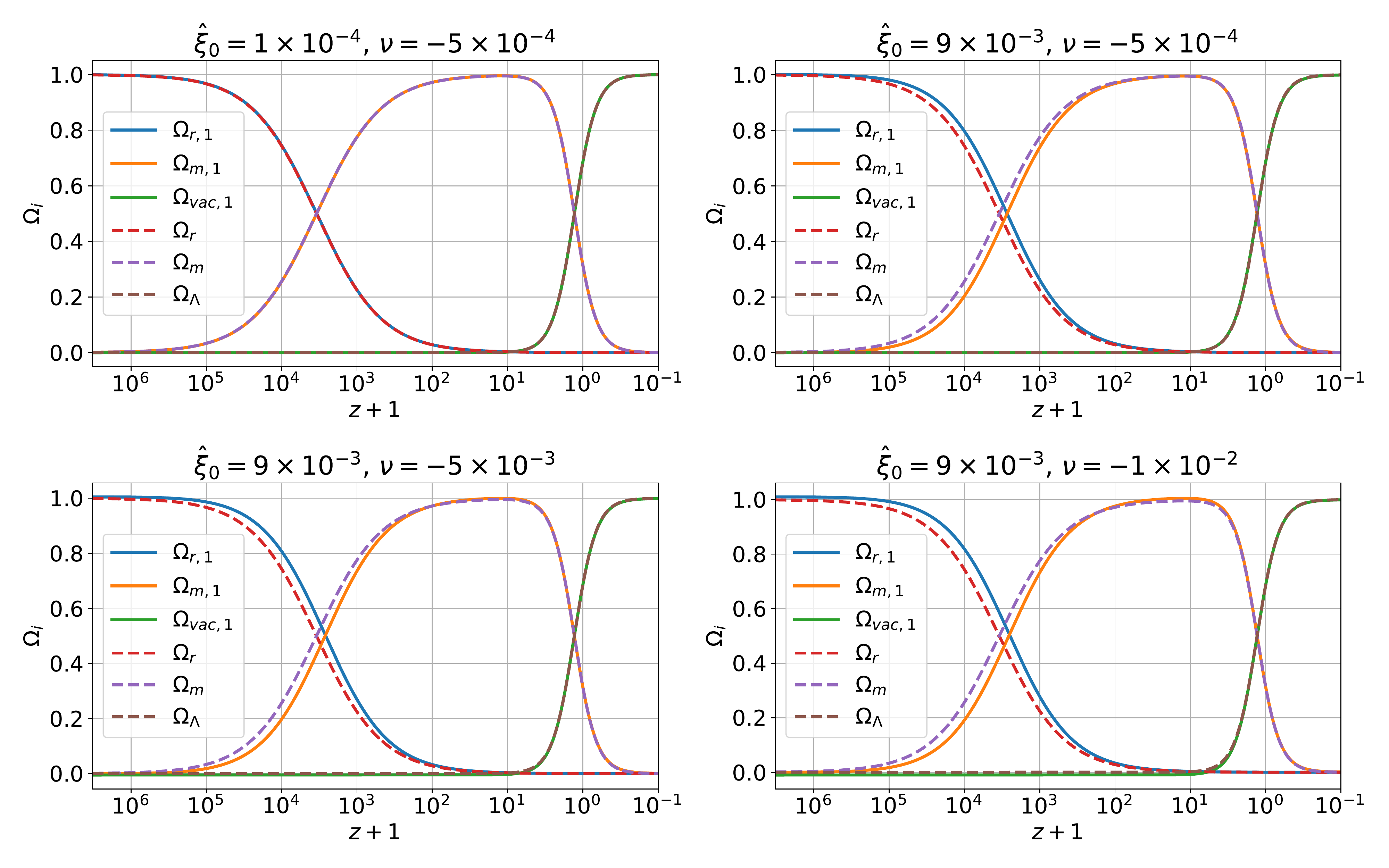}}
    \caption{\label{fig:M1DensityPlots} Plots of density parameters associated to each fluid $\Omega_{i,1}$ for \textbf{Model 1} as a function of redshift $z$, for different $\hat{\xi}_{0}$-values (solid lines). Positive and negative $\nu$-values are respectively considered in \textbf{(a)} and \textbf{(b)}. The dashed lines correspond to the density parameters $\Omega_{i}$ for $\Lambda$CDM model, obtained from Eq. \eqref{LCDM}, where $i$ stands for $r$ (radiation), $m$ (matter), and $\rm vac$ (vacuum). This model corresponds to the first class of models where $\tilde{\nu}=0$ with $s=1/2$, whose solutions are obtained by the numerical integration of Eq.~(\ref{sec3:eqn3}). The x-axis is presented in the $z+1$ range in order to obtain a better representation in the logarithm scale.}
\end{figure*}

\begin{figure*}
    \centering
    \subfigure[\label{fig:M1nu+Variationdensity}]{\includegraphics[scale = 0.35]{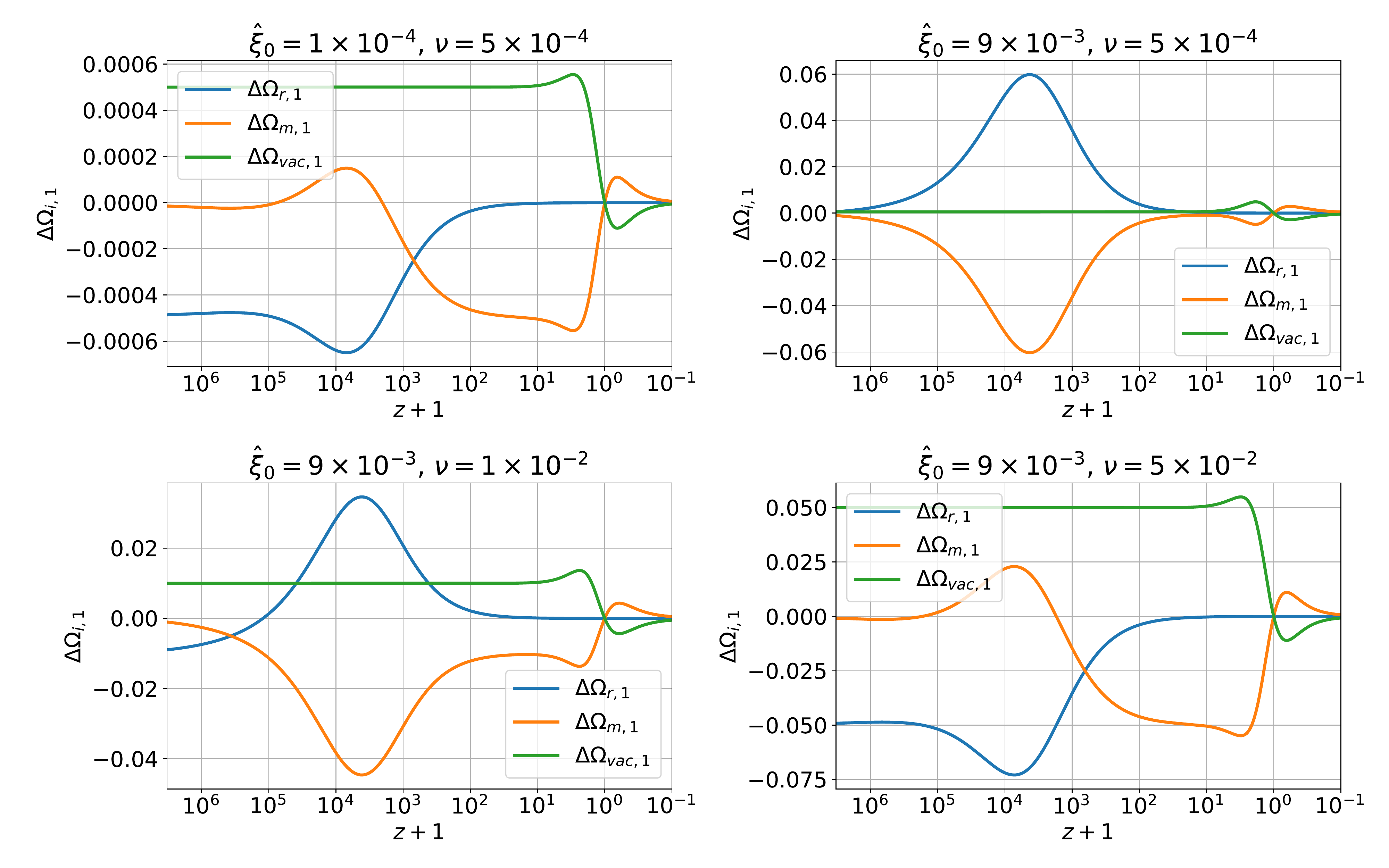}}
    \subfigure[\label{fig:M1nu-Variationdensity}]{\includegraphics[scale = 0.35]{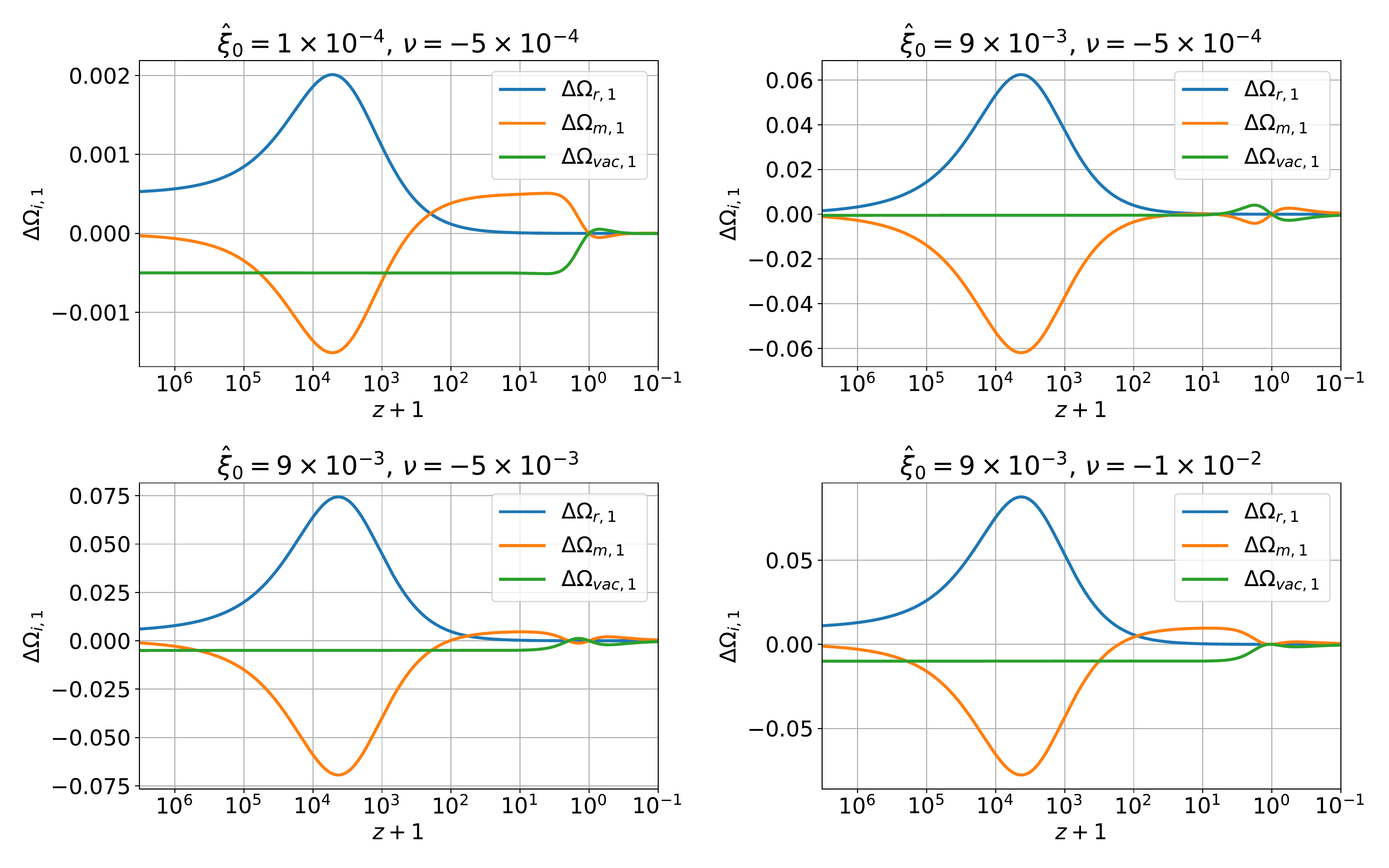}}
    \caption{\label{fig:M1VariationdensityPlots} Plots of the variation of the density parameters $\Delta\Omega_{i,1}$ associated to each fluid $\Omega_{i,1}$ for \textbf{Model 1}, with respect to their $\Lambda$CDM counterparts $\Omega_{i}$, as a function of redshift $z$, for different values of $\hat{\xi}_{0}$. Positive and negative $\nu$-values are respectively considered in \textbf{(a)} and \textbf{(b)}. The curves are obtained through the expression $\Delta\Omega_{i,1}=\Omega_{i,1}-\Omega_{i}$, where $i$ stands for $r$ (radiation), $m$ (matter), and $\rm vac$ or $\Lambda$ (vacuum).}
\end{figure*}

\begin{figure*}
    \centering
    \subfigure[\label{fig:M1nu+Barotropic}]{\includegraphics[scale = 0.35]{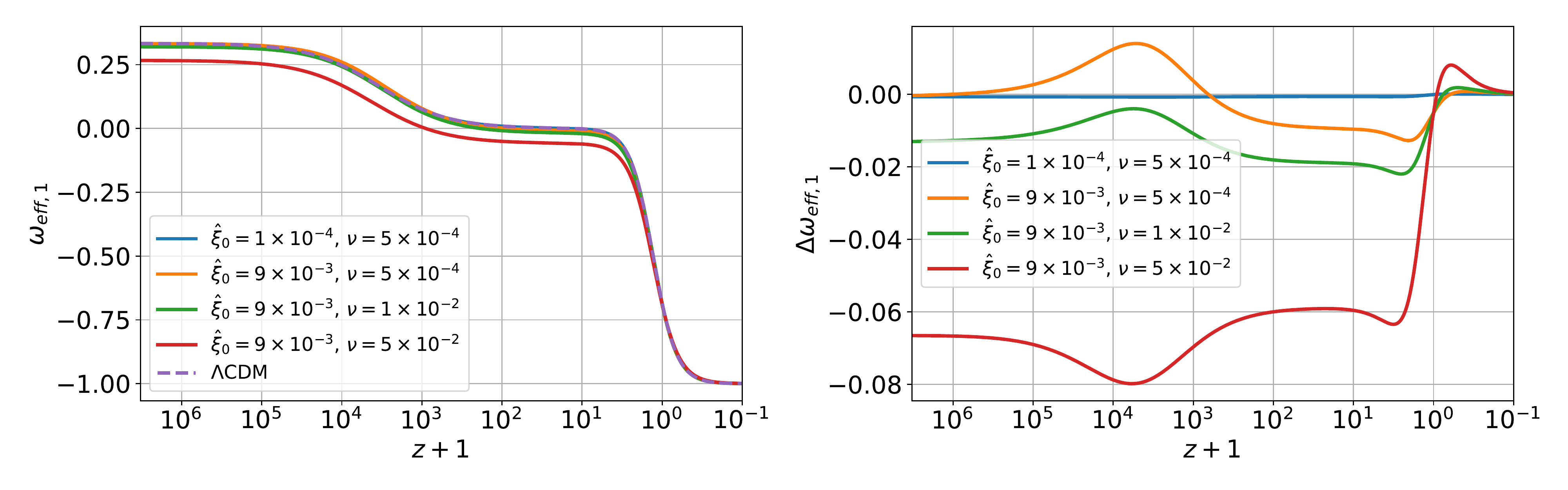}}
    \subfigure[\label{fig:M1nu-Barotropic}]{\includegraphics[scale = 0.35]{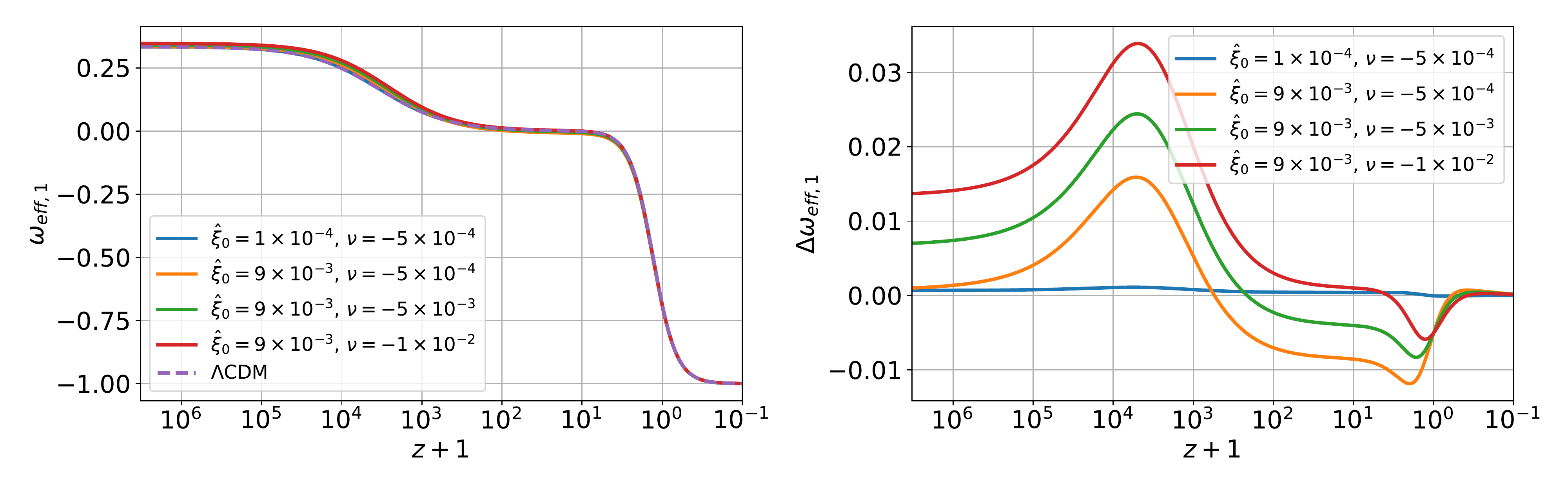}}
    \caption{\label{fig:M1BarotropicPlots} \textbf{(left)} Plot of the effective barotropic index $\omega_{\rm eff,1}$ for \textbf{Model 1}, obtained from Eq. \eqref{sec3:eqn5}, as a function of redshift $z$. \textbf{(right)}. Plot of the variation of the effective barotropic index $\Delta\omega_{\rm eff,1}=\omega_{\rm eff,1}-\omega_{\rm eff}$, were $\omega_{\rm eff}$ correspond to their $\Lambda$CDM counterpart obtained from Eq. \eqref{sec5:eqn2}, as a function of redshift $z$. Positive and negative $\nu$-values are respectively considered in \textbf{(a)} and \textbf{(b)}, for the  the same values of $\hat{\xi}_{0}$.}
\end{figure*}

\begin{figure*}
    \centering
    \subfigure[\label{fig:M1nu+Vacuum}]{\includegraphics[scale = 0.35]{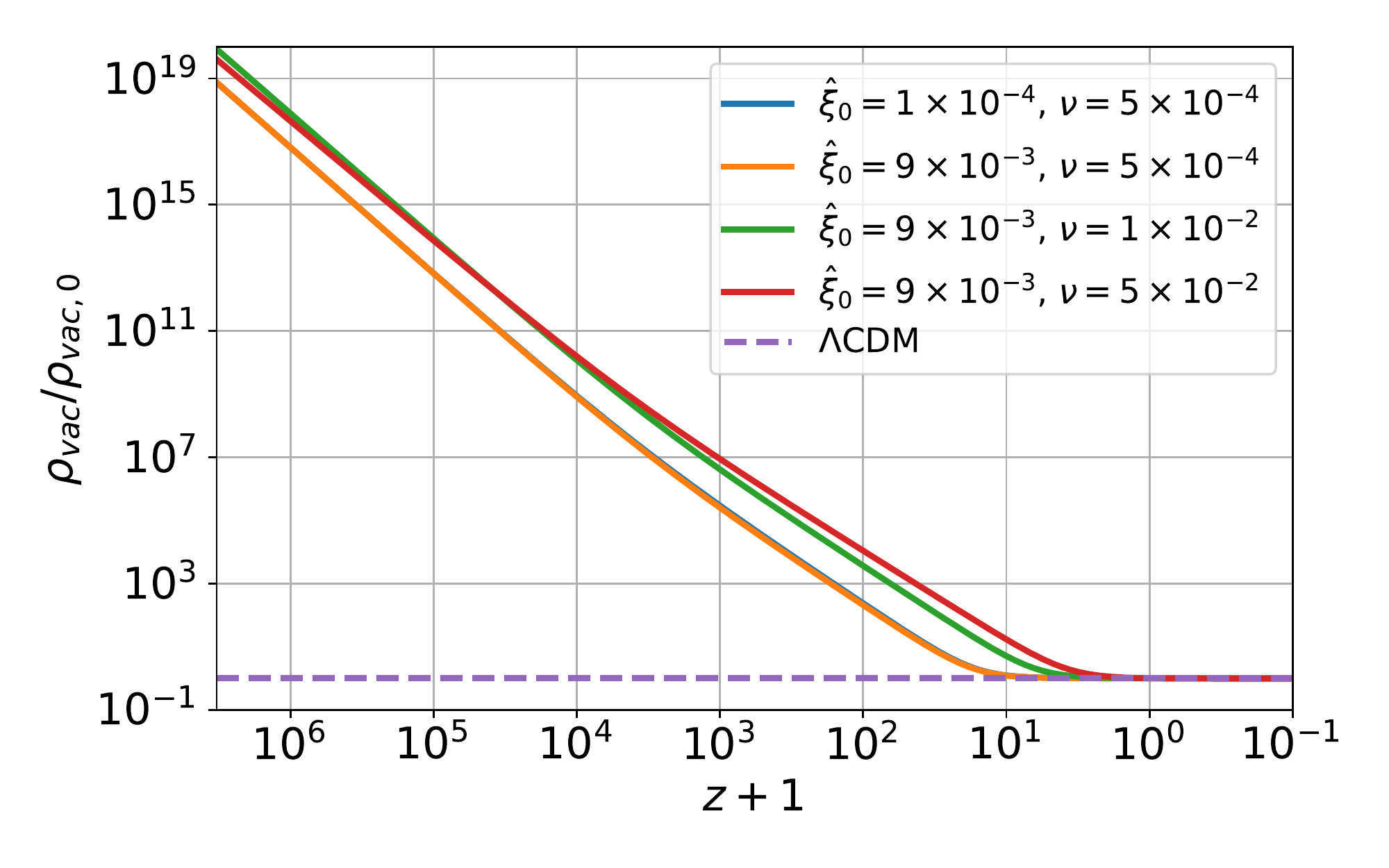}}
    \hspace{0.15cm}
    \subfigure[\label{fig:M1nu-Vacuum}]{\includegraphics[scale = 0.35]{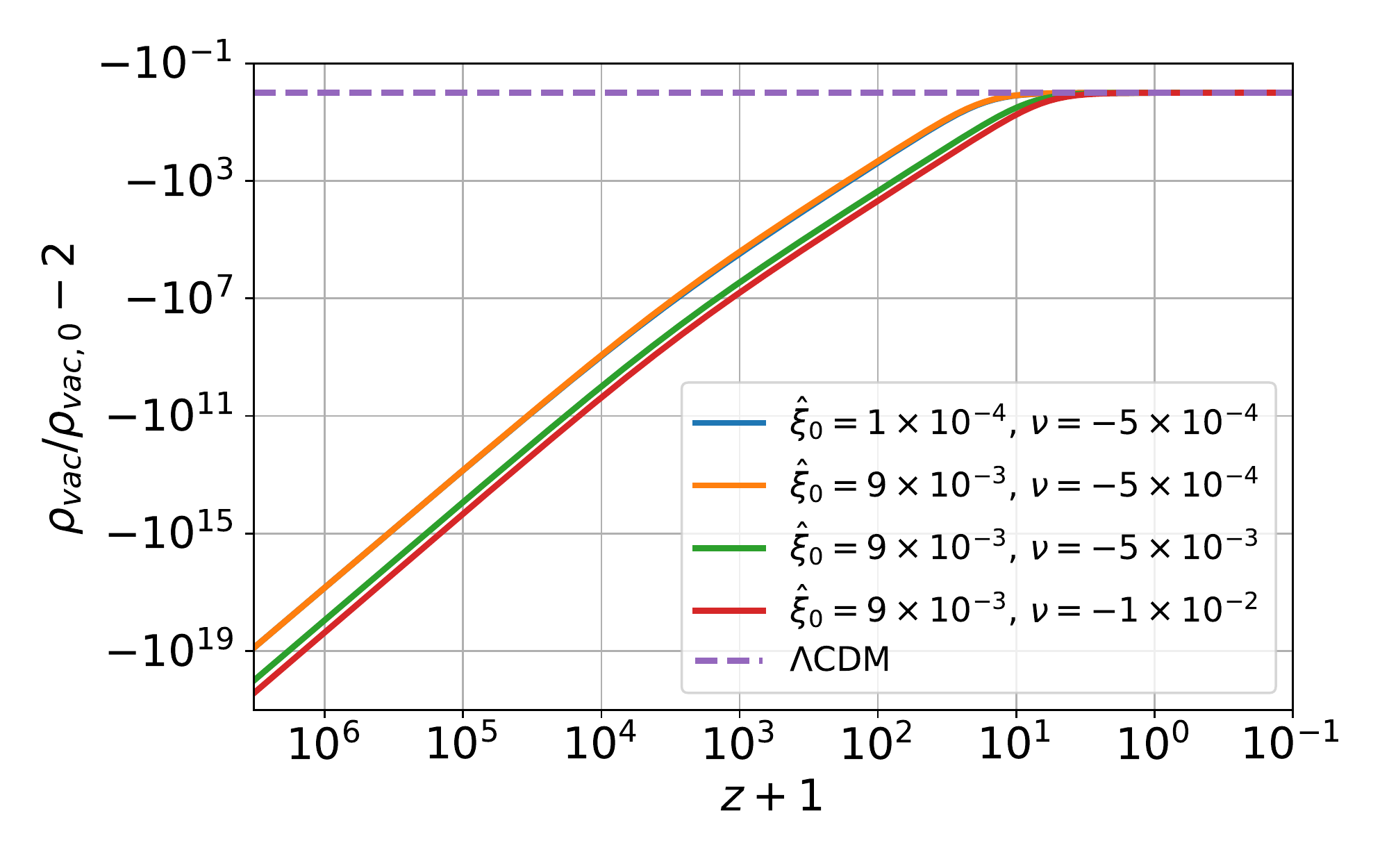}}
    \caption{\label{fig:M1Vacuum} Plots of vacuum energy density $\rho_{\rm vac}$ for \textbf{Model 1}, normalized with respect to their current value $\rho_{\rm vac,0}$, as a function of redshift $z$. Positive and negative $\nu$-values are respectively considered in \textbf{(a)} and \textbf{(b)}, for different $\hat{\xi}_{0}$-values. Notice that we have plotted $\rho_{\rm vac}/\rho_{\rm vac,0}-2$ in order to obtain a better representation in the symmetrical logarithm scale. As a reference, we have used $\Lambda=4.24\pm0.11\times 10^{-66}$ eV$^{2}$ \cite{Planck:2018vyg} to compute the present vacuum energy density $\rho_{\rm vac,0}$.} 
\end{figure*}

In figures \ref{fig:M1DensityPlots}, \ref{fig:M1VariationdensityPlots}, \ref{fig:M1BarotropicPlots}, and \ref{fig:M1Vacuum}, we present the numerical results for model 1, obtained by the integration of Eq. \eqref{sec3:eqn3} with $s=1/2$, and $\tilde{\nu}=0$.

In figure \ref{fig:M1DensityPlots}, we depict the energy density parameters $\Omega_{i,1}$ associated to each fluid component (i.e. $i$ goes from $r$ for radiation, $m$ for matter, to $\rm vac$ or $\Lambda$ for vacuum) as a function of redshift $z$, and for comparison their corresponding counterparts $\Omega_{i}$ for the $\Lambda$CDM model according to Eq. \eqref{LCDM}. In particular, Figure \ref{fig:M1nu+Density} shows the numerical results of the energy density parameters obtained for the different values of $\hat{\xi}_{0}$ and positive $\nu$, while in figure \ref{fig:M1nu-Density}  numerical results obtained for the different values of $\hat{\xi}_{0}$ and negative $\nu$ are presented. From these figures, we can see how the bulk viscosity and the running vacuum affect the redshift value at which the intersection $\Omega_{r,1}=\Omega_{m,1}$ happens (which we calle $z_{eq,1}$), without any appreciated effect in the redshift value at which $\Omega_{m,1}=\Omega_{\rm vac,1}$. Even more, in the case of $\nu>0$, it can be noted how the increment in the values of $\hat{\xi}_{0}$ implies that $z_{eq,1}<z_{eq}$, being $z_{eq}$ the redshift at which $\Omega_{r}=\Omega_{m}$; while the increment in the values of $\nu$ implies that $z_{eq}<z_{eq,1}$. On the contrary, in the case of $\nu<0$ the increment in the values of $\hat{\xi}_{0}$ and/or $\nu$ implies that $z_{eq,1}<z_{eq}$. It follows that it is possible to choose a combination of $\hat{\xi}_{0}$ and $\nu>0$ such that $z_{eq,1}=z_{eq}$.

In figure \ref{fig:M1VariationdensityPlots}, we depict the variation of the density parameters associated to each fluid component with respect to the $\Lambda$CDM model as a function of redshift $z$, according to the expression $\Delta\Omega_{i,1}=\Omega_{i,1}-\Omega_{i}$. In particular, numerical results of the variation of the density parameters obtained for the different values of $\hat{\xi}_{0}$ for positive $\nu$-values are displayed in figure  \ref{fig:M1nu+Variationdensity}, while in figure \ref{fig:M1nu-Variationdensity} negative $\nu$-values are considered. From these figures we can see (with greater details than what it is seen in the figure \ref{fig:M1DensityPlots}) how the bulk viscosity and the running vacuum affect the evolution of the density parameters $\Omega_{r,1}$, $\Omega_{m,1}$, and $\Omega_{\rm vac,1}$. In particular, it can be noted how a positive $\nu$ implies a larger value of $\Omega_{\rm vac,1}$ in comparison to $\Omega_{\Lambda}$; while a negative $\nu$ implies a smaller value of $\Omega_{\rm vac,1}$ in comparison to $\Omega_{\Lambda}$, which holds at high redshift. This is an expected result, as it can be seen from Eq. \eqref{sec2:eqn1}. On the other hand, the bulk viscosity and the running vacuum affect the evolution of $\Omega_{r,1}$ despite the fact that the bulk viscosity is associated to the matter and the running to the DE, i.e., radiation ``feels'' these effects because all the fluids are constrained trough Eq. \eqref{sec3:eqn2}. This analysis is in agreement with the critical points presented in the Table \ref{M1:critical points}, where we can see, for example, that one critical point for $\Omega_{r}$ is $1-\nu-9\hat{\xi}_{0}^{2}$ (in the plot the point is $-\nu-9\hat{\xi}_{0}^{2}$). Due to the small values of $\hat{\xi}_{0}$, the effects of the bulk viscosity are visible between the current time and a high redshift, and even more notable for $z+1\approx 10^{3}-10^{4}$. It is important to note that the vacuum density parameter seems not negligible at a very high redshift, with an apparent constant behavior for $z+1>10$ in both cases.

In figure \ref{fig:M1BarotropicPlots}, we depict the effective barotropic index $\omega_{\rm eff,1}$, according to Eq. \eqref{barotropic_const}, and its deviation from the effective barotropic index $\omega_{\rm eff}$ of $\Lambda$CDM model obtained from Eq. \eqref{sec5:eqn2}, which is defined by $\Delta\omega_{\rm eff,1}=\omega_{\rm eff,1}-\omega_{\rm eff}$. In particular, in figure \ref{fig:M1nu+Barotropic} the effective barotropic index and the difference $\Delta\omega_{\rm eff,1}$ are shown as a function of redshift for different $\hat{\xi}_{0}$ values and $\nu >0$. For comparison, the corresponding quantity for the standard $\Lambda$CDM is displayed. The same representation is shown in figure \ref{fig:M1nu-Barotropic} for different $\hat{\xi}_{0}$ values and $\nu<0$. From these figures we can see how the bulk viscosity and the running vacuum affect the evolution of $\omega_{\rm eff,1}$, being remarkably different when $|\nu|$ take larger values. Nevertheless, this behavior is a consequence of the small size of $\hat{\xi}_{0}$, since there are appreciated effects for larger values of this parameter. Focusing in the effects due to the sign of $\nu$, we can see that for $\nu>0$ the values of $\omega_{\rm eff,1}$ are lower than $\omega_{\rm eff}$ at high redshift, which is a consequence of a positive not negligible $\Omega_{\rm vac,1}$; while for a $\nu<0$ the values of $\omega_{\rm eff,1}$ are greater than $\omega_{\rm eff}$ at high redshift, which is a consequence of a negative not negligible $\Omega_{\rm vac,1}$. At low redshift there is a change of this behaviour for the $\nu<0$ case. It is important to mention that the possibility of $\Omega_{\rm vac,1}<0$ comes from the definition of $\rho_{\rm vac}$ as is discussed below.

In figure \ref{fig:M1Vacuum}, we depict the vacuum energy density normalized with respect to their current value as a function of the redshift $z$, as well as the normalized vacuum energy density for the $\Lambda$CDM model for a further comparison. In figure \ref{fig:M1nu+Vacuum} the normalized vacuum energy density is displayed for the different values of $\hat{\xi}_{0}$ and positive $\nu$, while negative $\nu$ values are presented in figure \ref{fig:M1nu-Vacuum}. From this figures we can see how the bulk viscosity and the running vacuum affect the evolution of the vacuum energy density. It follows that an increment in $\hat{\xi}_{0}$ does not appreciably affect the evolution of $\rho_{\rm vac}$, contrary to what happens when we increment the values of $|\nu|$. This behaviour is due to the fact that the bulk viscosity affects the evolution of $\rho_{\rm vac}$ indirectly through the Hubble parameter according to the Eq. \eqref{sec2:eqn1} and, therefore, it is necessary a remarkably difference in the evolution of $H$ that does not appear due to the small values of $\hat{\xi}_{0}$. On the other hand, depending on the sign of $\nu$, it is possible to obtain an always positive vacuum energy density when $\nu>0$ or a vacuum energy density that experiences a transition between positive to negative values when $\nu<0$. From Eq. \eqref{sec2:eqn1}, this transition occurs when
\begin{equation}\label{transitionnotdotH}
    H=H_{0}\sqrt{\frac{\Omega_{\rm vac,0}+|\nu|}{|\nu|}},
\end{equation}
and therefore, considering that $H=H(z)$, the redshift at which this change of sign occur depends strongly on the values of $\nu$. It is worthwhile pointing out that the contribution of the running vacuum could reach at very high redshift large values of the order $10^{19}$ with respect to its current value, which according to its measured value is $\Lambda=4.24\pm0.11\times 10^{-66}$ eV$^{2}$ \cite{Planck:2018vyg}. Despite that the effective vacuum energy density would be still small, the Hubble parameter would become a very large value. This leads to the question whether large redshift values such as $z = 10^6$ lie within the validity range of the Ansatz given by Eq.\eqref{sec2:eqn1}. The answer to this task amounts including further terms of order $H^4$ in the expansion of running vacuum energy, and therefore goes beyond the scope of the present work (for a discussion of different running vacuum energy expansions see ref. \cite{Sola:2015rra}).    

\subsection{\label{NUmericalM2} Model 2}
\begin{figure*}
    \centering
    \subfigure[\label{fig:M2nu+Density}]{\includegraphics[scale = 0.35]{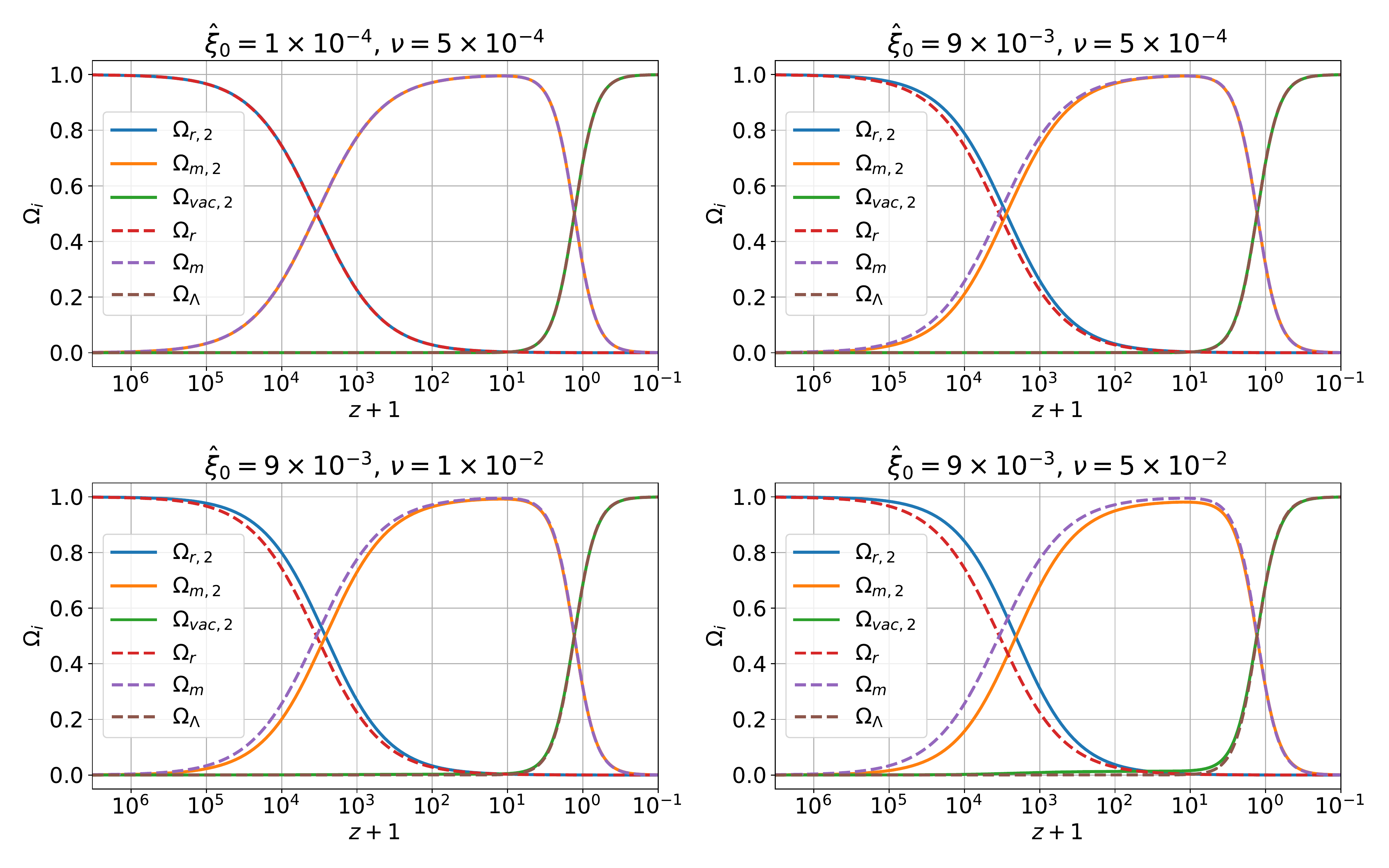}}
    \subfigure[\label{fig:M2nu-Density}]{\includegraphics[scale = 0.35]{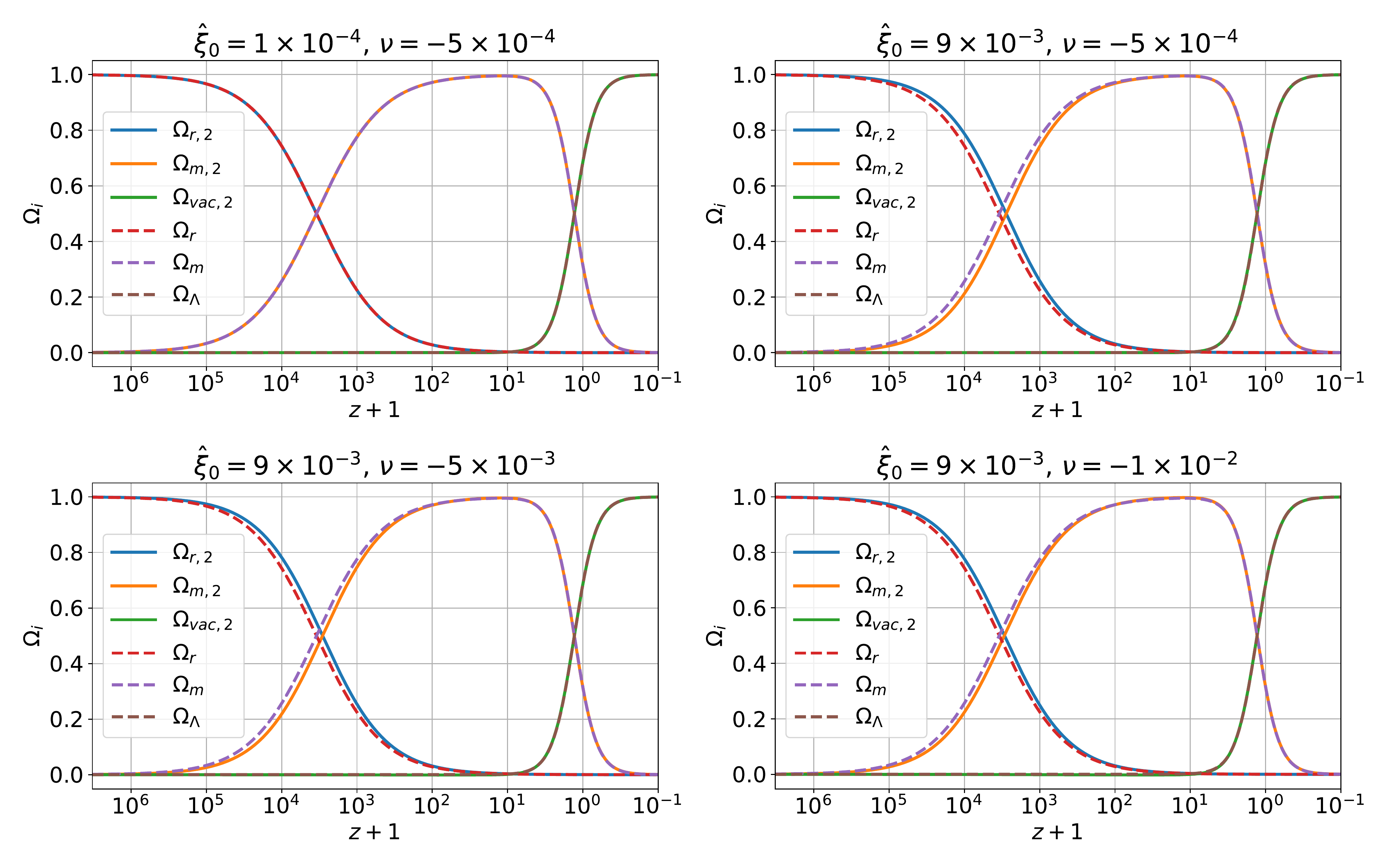}}
    \caption{\label{fig:M2DensityPlots} Plots of density parameters associated to each fluid $\Omega_{i,2}$ for \textbf{Model 2} as a function of redshift $z$, for different values of $\hat{\xi}_{0}$ (solid lines). Positive and negative $\nu$-values are respectively considered in \textbf{(a)} and \textbf{(b)}. The dashed lines correspond to the density parameters $\Omega_{i}$ for the $\Lambda$CDM model, obtained from Eq. \eqref{LCDM}, where $i$ stands for $r$ (radiation), $m$ (matter), and $\rm vac$ (vacuum). This model corresponds to the second class of models where $\tilde{\nu}=\nu/2$ with $s=1$, whose solutions are obtained by the numerical integration of Eq.~(\ref{sec3:eqn4}).}
\end{figure*}

\begin{figure*}
    \centering
    \subfigure[\label{fig:M2nu+Variationdensity}]{\includegraphics[scale = 0.35]{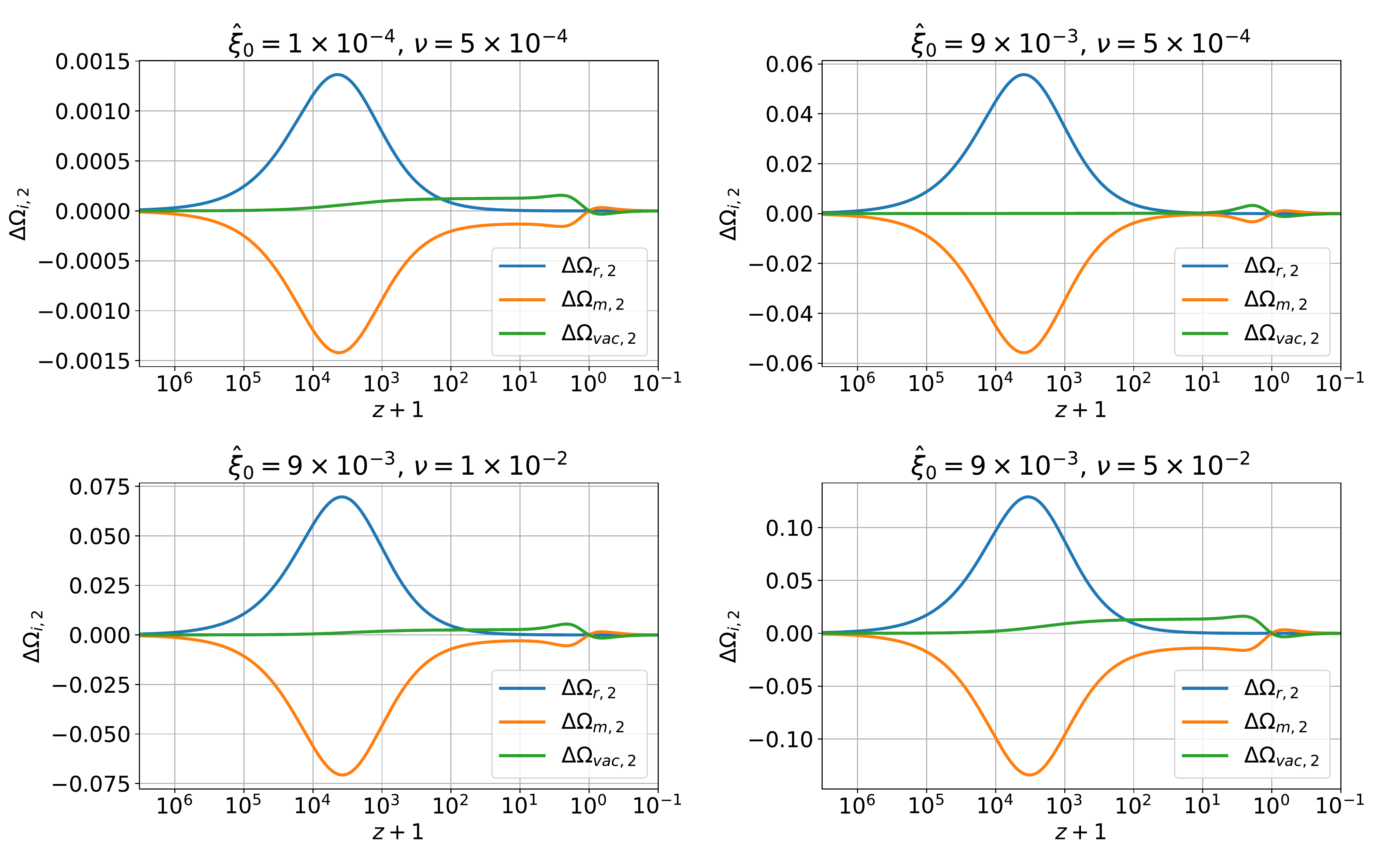}}
    \subfigure[\label{fig:M2nu-Variationdensity}]{\includegraphics[scale = 0.35]{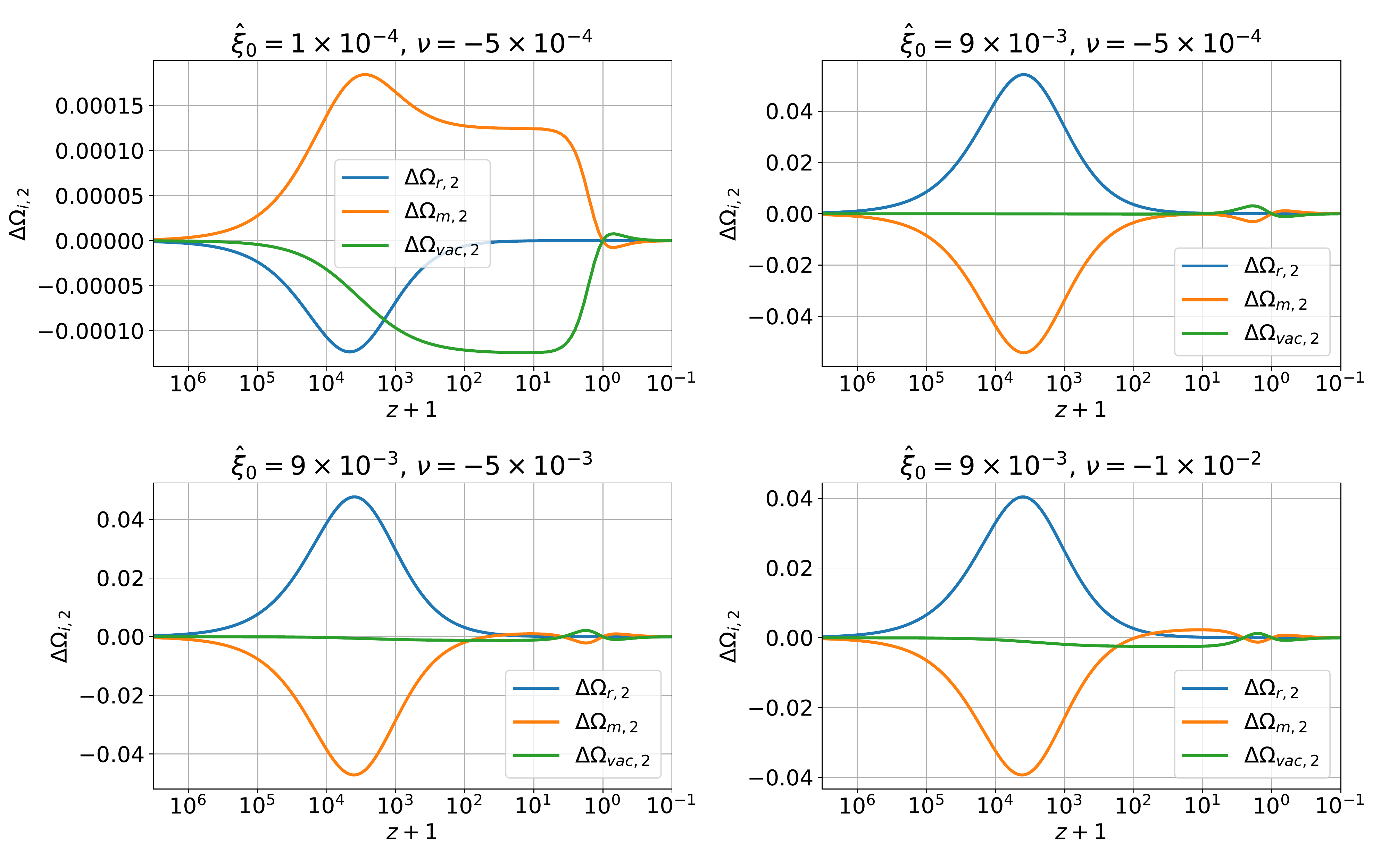}}
    \caption{\label{fig:M2VariationdensityPlots} Plots of the variation of the density parameters $\Delta\Omega_{i,2}$ associated to each fluid $\Omega_{i,2}$ for \textbf{Model 2}, with respect to their $\Lambda$CDM counterparts $\Omega_{i}$, as a function of redshift $z$, for different values of $\hat{\xi}_{0}$. Positive and negative $\nu$-values are respectively considered in \textbf{(a)} and \textbf{(b)}. The curves are obtained through the expression $\Delta\Omega_{i,2}=\Omega_{i,2}-\Omega_{i}$, where $i$ stands for $r$ (radiation), $m$ (matter), and $\rm vac$ (vacuum).}
\end{figure*}

\begin{figure*}
    \centering
    \subfigure[\label{fig:M2nu+Barotropic}]{\includegraphics[scale = 0.35]{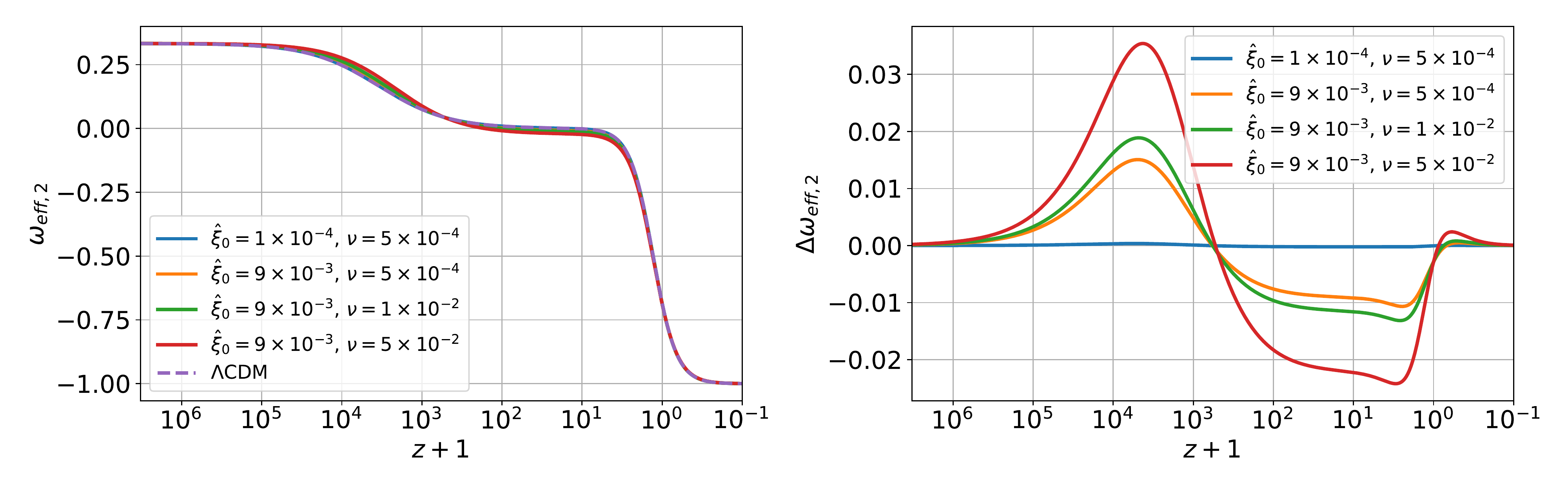}}
    \subfigure[\label{fig:M2nu-Barotropic}]{\includegraphics[scale = 0.35]{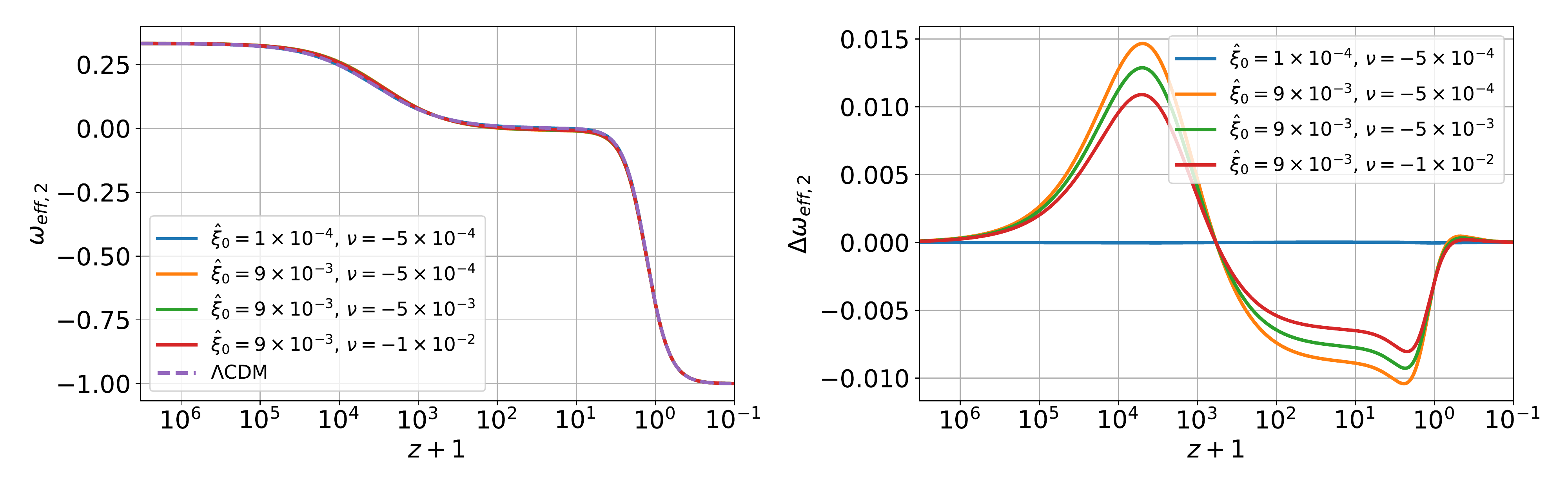}}
    \caption{\label{fig:M2BarotropicPlots} \textbf{(left)} Plot of the effective barotropic index $\omega_{\rm eff,2}$ for \textbf{Model 2}, obtained from Eq. \eqref{sec3:eqn5}, as a function of redshift $z$. \textbf{(right)}. Plot of the variation of the effective barotropic index $\Delta\omega_{\rm eff,2}=\omega_{\rm eff,2}-\omega_{\rm eff}$, were $\omega_{\rm eff}$ correspond to their $\Lambda$CDM counterpart obtained from Eq. \eqref{sec5:eqn2}, as a function of redshift $z$. Positive and negative $\nu$-values are respectively considered in \textbf{(a)} and \textbf{(b)}, for the same values of $\hat{\xi}_{0}$.}
\end{figure*}

\begin{figure*}
    \centering
    \subfigure[\label{fig:M2nu+Vacuum}]{\includegraphics[scale = 0.35]{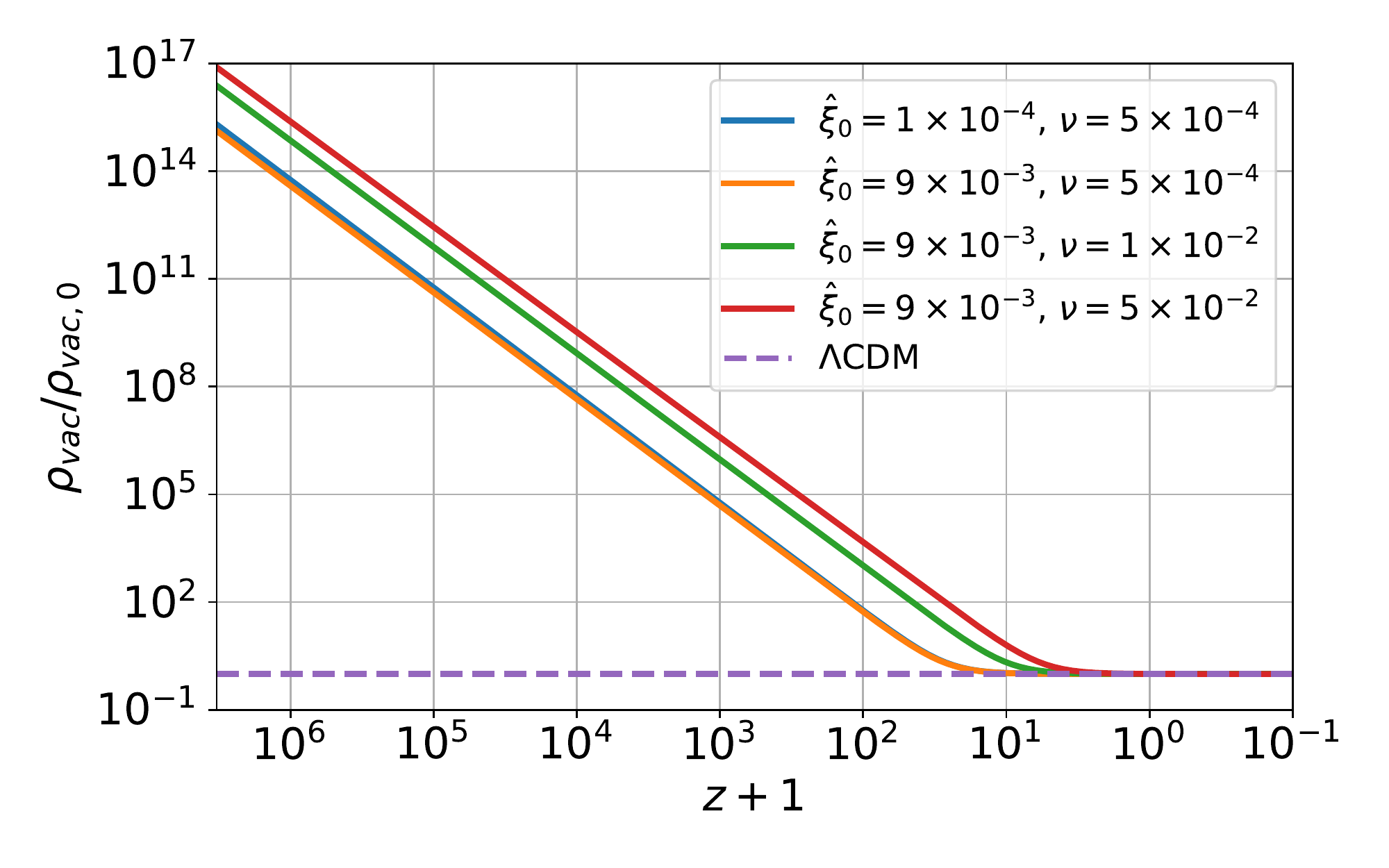}}
    \hspace{0.15cm}
    \subfigure[\label{fig:M2nu-Vacuum}]{\includegraphics[scale = 0.35]{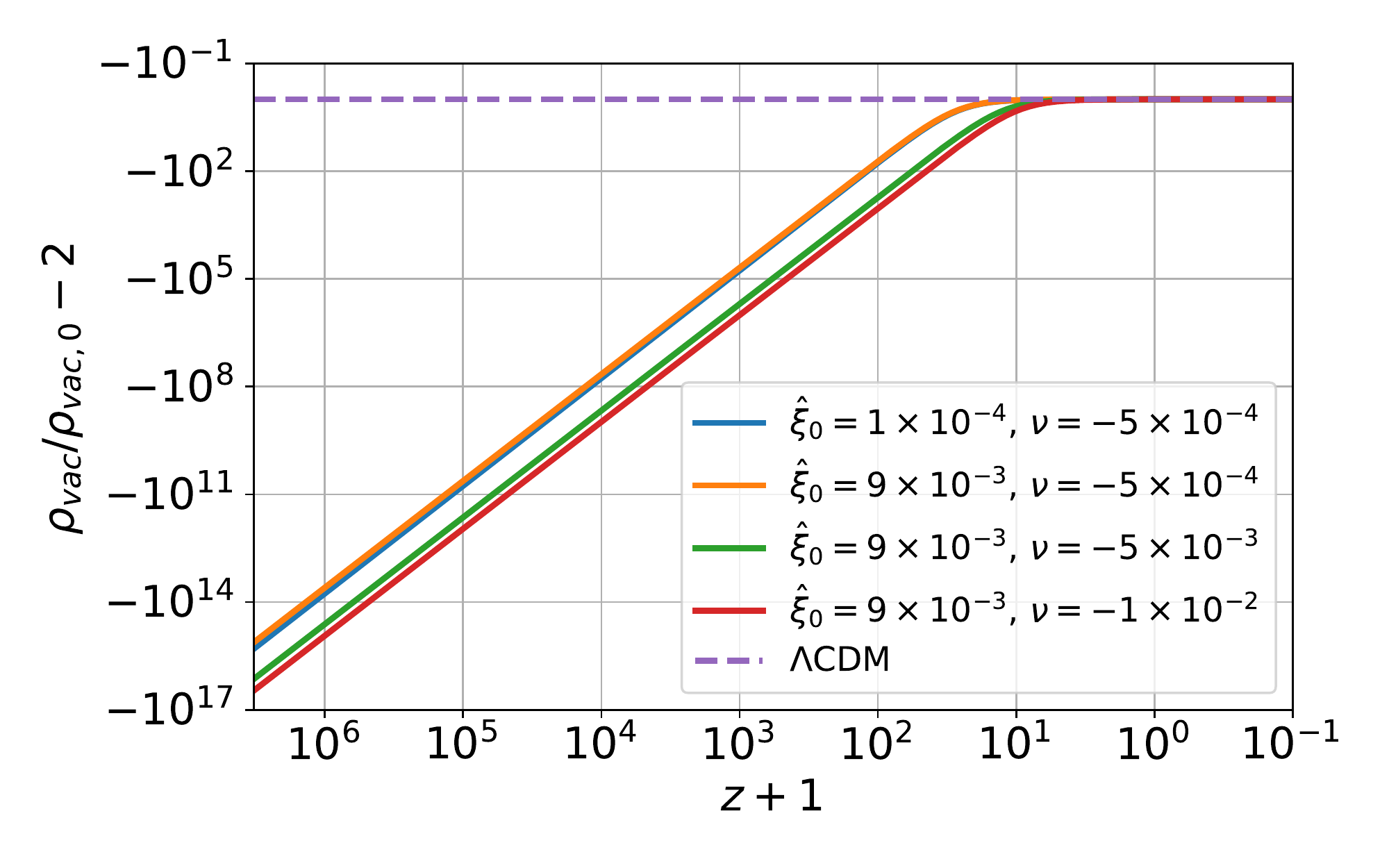}}
    \caption{\label{fig:M2Vacuum} Plots of vacuum energy density $\rho_{\rm vac}$ for \textbf{Model 2}, normalized with respect to their current value $\rho_{\rm vac,0}$, as a function of redshift $z$. Positive and negative $\nu$-values are respectively considered in \textbf{(a)} and \textbf{(b)}, for different $\hat{\xi}_{0}$-values. Notice that we have plotted $\rho_{\rm vac}/\rho_{\rm vac,0}-2$ in order to obtain a better representation in the symmetrical logarithm scale. As a reference, we have used $\Lambda=4.24\pm0.11\times 10^{-66}$ eV$^{2}$ \cite{Planck:2018vyg} to compute the present vacuum energy density $\rho_{\rm vac,0}$.}
\end{figure*}

In figures \ref{fig:M2DensityPlots}, \ref{fig:M2VariationdensityPlots}, \ref{fig:M2BarotropicPlots}, and \ref{fig:M2Vacuum}, we present the numerical results for model 2, obtained by the integration of Eq. \eqref{sec3:eqn4} with $s=1$. As a reminder, this model corresponds to the second class of models with $\tilde{\nu}=\nu/2$. 

In figure \ref{fig:M2DensityPlots}, we show the density parameters $\Omega_{i,2}$ associated to each fluid component as a function of redshift $z$, as well as the counterparts density parameters $\Omega_{i}$ of $\Lambda$CDM model, according to Eq. \eqref{LCDM}, aiming to a further comparison. In figure \ref{fig:M2nu+Density} we show numerical results for the density parameters obtained for different values of $\hat{\xi}_{0}$ and positive $\nu$, while in figure \ref{fig:M2nu-Density} negative $\nu$ are considered. From these figures we can see how the bulk viscosity and the running vacuum affect the redshift value (which we call $z_{eq,2}$) at which $\Omega_{r,2}=\Omega_{m,2}$, while the other intersection $\Omega_{m,2}=\Omega_{\rm vac,2}$ remains essentially at the same redshift value, independent of dissipation and running. In this sense, and contrarily to what happens in the model 1, both positive and negative $\nu$ lead to $z_{eq,2}<z_{eq}$, being $z_{eq}$ the redshift value at which $\Omega_{r}=\Omega_{m}$ for $\Lambda$CDM model. Nevertheless, and as a comparison to what happens in model 1, the values of $z_{eq,2}$ are closer to $z_{eq}$ for all $\nu<0$. This is a consequence of the behaviour of $\rho_{\rm vac}$ as we will argue below.

In figure \ref{fig:M2VariationdensityPlots}, we depict the differences $\Delta\Omega_{i,2}=\Omega_{i,2}-\Omega_{i}$ of the density parameters associated to each fluid component with respect to the corresponding ones in the $\Lambda$CDM model as a function of redshift $z$. In figure \ref{fig:M2nu+Variationdensity} we present the numerical results of the variation of the density parameters obtained for different values of $\hat{\xi}_{0}$ and positive $\nu$, while in figure \ref{fig:M2nu-Variationdensity} we present corresponding results but with negative $\nu$. From these figures we can see (with better detail than what it is seen in figure \ref{fig:M2DensityPlots}) how the bulk viscosity and the running vacuum affect the evolution of the density parameters $\Omega_{r,2}$, $\Omega_{m,2}$, and $\Omega_{\rm vac,2}$. Following this line, can be noted how a positive $\nu$ implies a greater value of $\Omega_{\rm vac,2}$ in comparison to $\Omega_{\Lambda}$; while a negative $\nu$ implies a lower value of $\Omega_{\rm vac,2}$ in comparison to $\Omega_{\Lambda}$. But, contrary to what happens in the model 1, this differences occurs only at low redshift because at very high redshift there is not remarkably differences in the three fluids with respect to the $\Lambda$CDM model. This is an expected result, as can be seen from Eq. \eqref{sec2:eqn1}, because in this model we add to the vacuum energy density an extra contribution that depends on $\dot{H}=-H(1+z)dH/dz$, which is a negative contribution for $H>0$, considering that $dH/dz$ is positive as $z$ grows. Therefore, at very high redshift only affect the evolution of the fluids the bulk viscous constant $\hat{\xi}_{0}$, whose contribution is negligible due to its small values, but visible between the current time and a high redshift, most notable in $z+1\approx 10^{3}-10^{4}$. This analysis is in agreement with the critical points presented in the Table \ref{M2:critical points}, where we can see, for example, that one critical point for $\Omega_{r}$ is $1$ (in the plot the point is $0$). It is important to mention that the difference between $\Omega_{\rm vac,2}$ and $\Omega_{\Lambda}$ are lower for $\nu<0$ because in this case, following Eq. \eqref{sec2:eqn1}, the contribution of the $\dot{H}$ term in the expansion is positive and the contribution of the $H^{2}$ term is negative. This last one analyses indicate that the major contribution to $\rho_{\rm vac}$ comes from the $H^{2}$ term than the $\dot{H}$ term.

In figure \ref{fig:M2BarotropicPlots}, we depict the effective barotropic index $\omega_{\rm eff,2}$, according to Eq. \eqref{sec3:eqn5}, and its deviation with respect to the effective barotropic index $\omega_{\rm eff}$ of the $\Lambda$CDM model, obtained from Eq. \eqref{sec5:eqn2}, which is defined by $\Delta\omega_{\rm eff,2}=\omega_{\rm eff,2}-\omega_{\rm eff}$. In particular, in figure \ref{fig:M2nu+Barotropic} the effective barotropic index and the difference $\Delta\omega_{\rm eff,2}$ are shown as a function of redshift for different $\hat{\xi}_{0}$ values and $\nu>0$. For comparison, the corresponding quantity for the standard $\Lambda$CDM model is displayed. The same representation are shown in figure \ref{fig:M2nu-Barotropic} for different $\hat{\xi}_{0}$ values and $\nu<0$. From these figures we can see how the bulk viscosity and the running vacuum affect the evolution of $\omega_{\rm eff,2}$, being remarkably different when $|\nu|$ take larger values. Nevertheless, this behavior is a consequence of the small size of $\hat{\xi}_{0}$, since there are appreciated effects for greater values of this parameter. Focusing in the effects of the sign of $\nu$, we can see that for both, positive and negative, the values of $\omega_{\rm eff,2}$ are greater than $\omega_{\rm eff}$ at high redshift, with a change of this behaviour at low redshift (similar to what happens in the model 1 for $\nu<0$). Even more, and contrary to whats happens in the model 1, at very high redshift there are no differences between $\omega_{\rm eff,2}$ and $\omega_{\rm eff}$, again due to negligible behaviour of $\Omega_{\rm vac,2}$ at this redshift.

In figure \ref{fig:M2Vacuum}, we depict the vacuum energy density normalized with respect to their current value as a function of redshift $z$, as well as the normalized vacuum energy density for $\Lambda$CDM model for a further comparison. In figure \ref{fig:M2nu+Vacuum} the normalized vacuum energy density is displayed for different values of $\hat{\xi}_{0}$ and positive $\nu$, while negative $\nu$ values are presented in figure \ref{fig:M2nu-Vacuum}. From this figures we can see how the bulk viscosity and the running vacuum affect the evolution of the vacuum energy density. In this sense an increment in $\hat{\xi}_{0}$ does not appreciably affect the evolution of $\rho_{\rm vac}$, contrary to what happens whith an increment in the values of $|\nu|$ (but note that this effect is not negligible). This behaviour is due to the fact that the bulk viscosity affects the evolution of $\rho_{\rm vac}$ through the Hubble parameter and its time derivative according to the Eq. \eqref{sec2:eqn1} and, therefore, it is necessary a remarkably difference in the evolution of $H$ that does not appear due to the small values of $\hat{\xi}_{0}$. On the other hand, depending on the sign of $\nu$ it is possible to obtain an always positive vacuum energy density when $\nu>0$ or a vacuum energy density that experience a transition between a positive to a negative one when $\nu<0$. From Eq. \eqref{sec2:eqn1}, this transition occurs when
\begin{equation}\label{transitiondotH}
    H^{2}=C(1+z)+\frac{H_{0}^{2}\left[\Omega_{\rm vac,0}+\frac{|\nu|}{2}(1-q_{0})\right]}{|\nu|},
\end{equation}
where $q_{0}$ is the current value of the deceleration parameter, related to the Hubble parameter through the expression $\dot{H}/H^{2}=-(1+q)$, and $C$ is an integration constant. The above equation depends explicitly in the r.h.s on the redshift, contrary to whats happens in Eq. \eqref{transitionnotdotH} which depends in their r.h.s only on constant terms. Therefore, it is necessary to know the dependency in $z$ of $H$ in order to obtain the redshift in which the change of sing in the vacuum energy density occurs. This kind of behavior, where a dynamical vacuum energy density can takes negative values at a finite redshift has been considered to alleviated low-redshift tensions, including the $H_{0}$ tension,\cite{akarsu2020graduated, banihashemi2019ginzburg,dutta2020beyond,Ye:2020btb}. It is interesting to note that this change of sing also occurs in other approaches, as for example the graduated dark energy, which phenomenologically describes a cosmological constant whose sign changes at a certain redshift, becoming positive just in the late time evolution \cite{akarsu2020graduated}. It is important to mention that the maximal contribution of the running vacuum energy density reaches $10^{17}$ at high redshift ($z =10^6$), which is two order of magnitude smaller than the one obtained for the model 1. Still, as the maximal value for the Hubble parameter becomes very large, the RG-inspired Ansatz of Eq.\eqref{sec2:eqn1} may go beyond its validity range. This issue cannot be addressed without the inclusion of further $H$-power contributions, which again goes beyond this present work.

\subsection{\label{NUmericalM3} Model 3}
\begin{figure*}
    \centering
    \subfigure[\label{fig:M3nu+Density}]{\includegraphics[scale = 0.35]{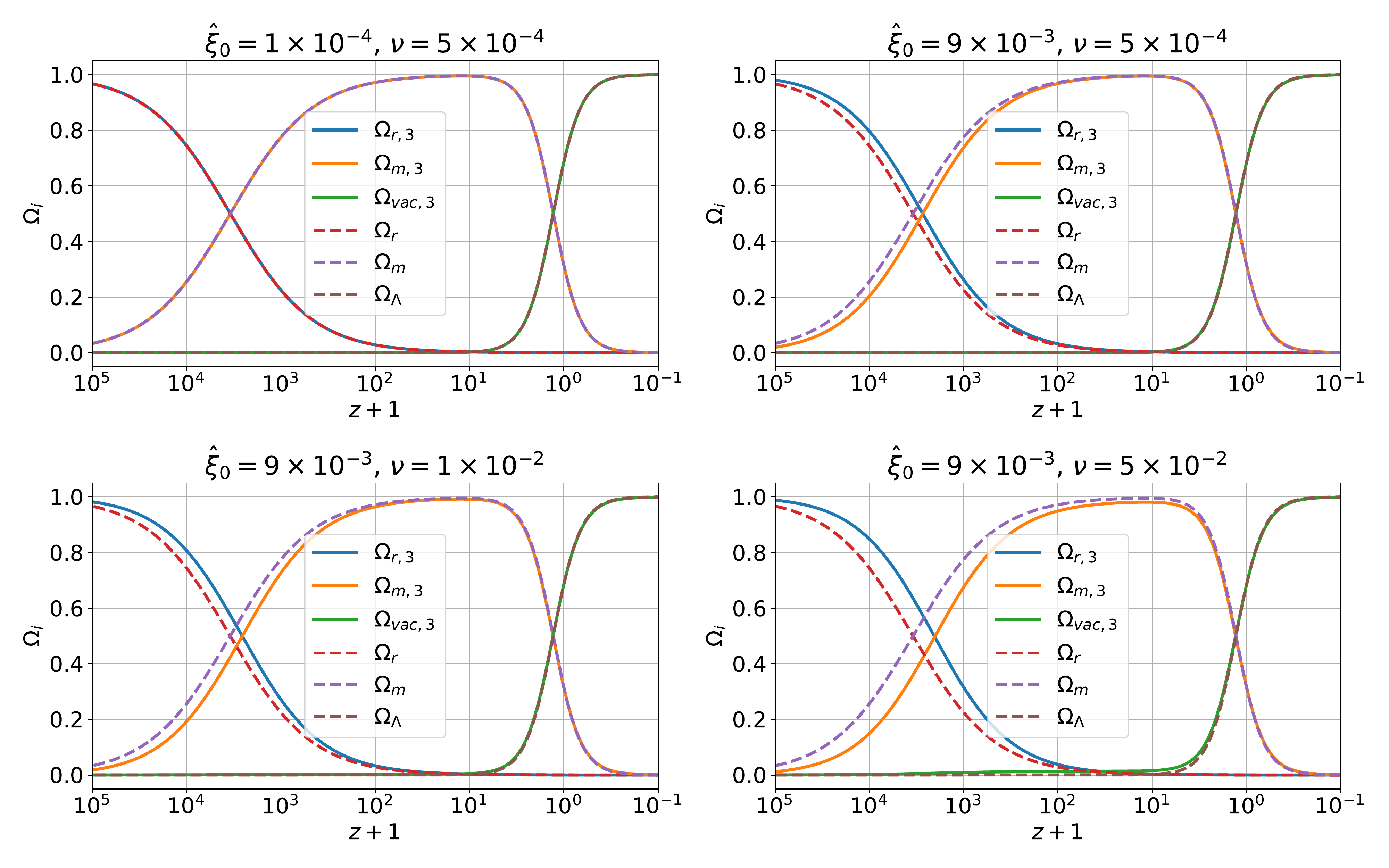}}
    \subfigure[\label{fig:M3nu-Density}]{\includegraphics[scale = 0.35]{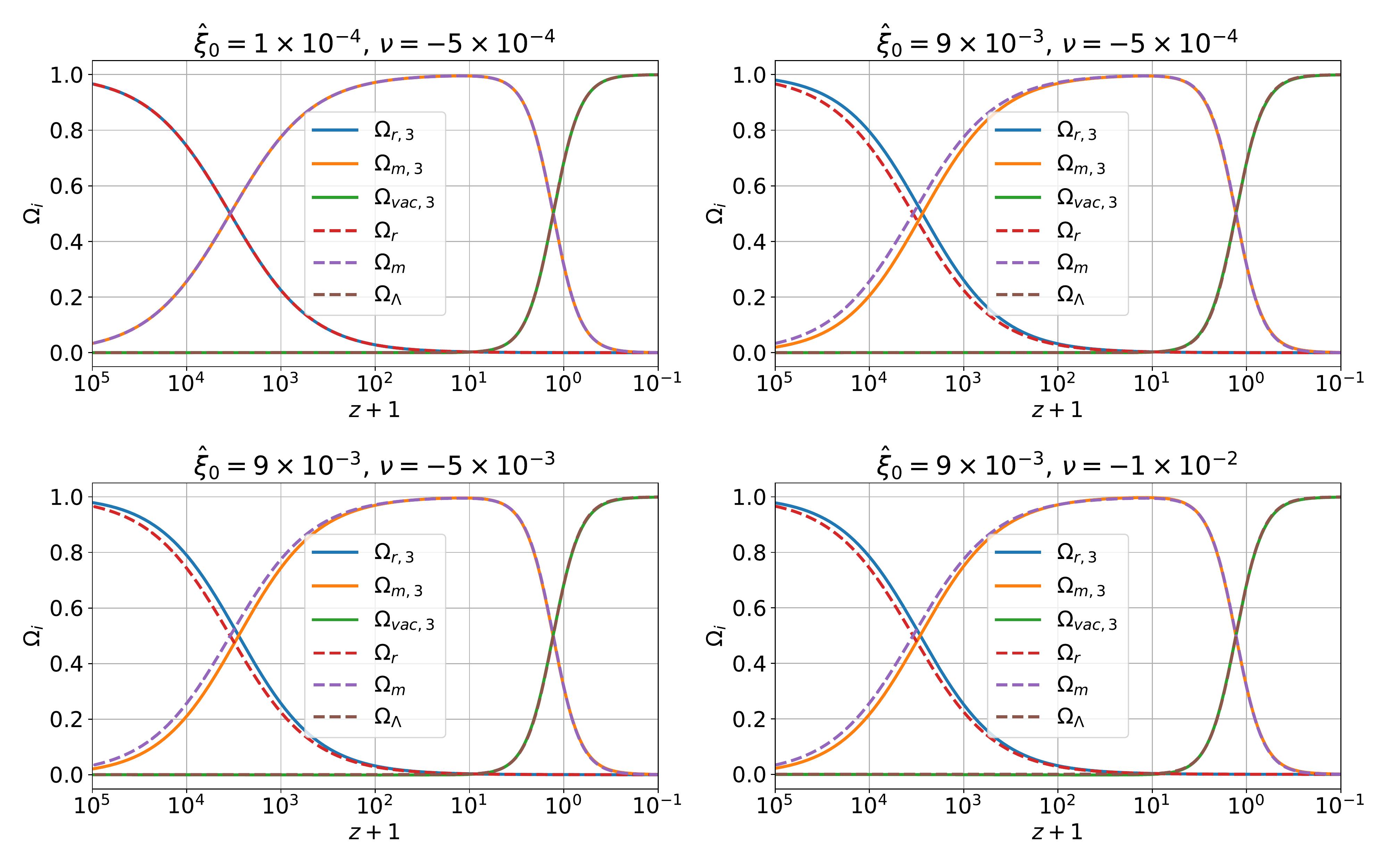}}
    \caption{\label{fig:M3DensityPlots} Plots of density parameters associated to each fluid $\Omega_{i,3}$ for \textbf{Model 3} as a function of redshift $z$, for different values of $\hat{\xi}_{0}$ (solid lines). Positive and negative $\nu$-values are respectively considered in \textbf{(a)} and \textbf{(b)}. The dashed lines correspond to the density parameters $\Omega_{i}$ for the $\Lambda$CDM model, obtained from Eq. \eqref{LCDM}, where $i$ stands for $r$ (radiation), $m$ (matter), and $\rm vac$ (vacuum). This model corresponds to the second class of models where $\tilde{\nu}=\nu/2$ with $s=1/2$, whose solutions are obtained by the numerical integration of Eq.~(\ref{sec3:eqn4}).}
\end{figure*}

\begin{figure*}
    \centering
    \subfigure[\label{fig:M3nu+Variationdensity}]{\includegraphics[scale = 0.35]{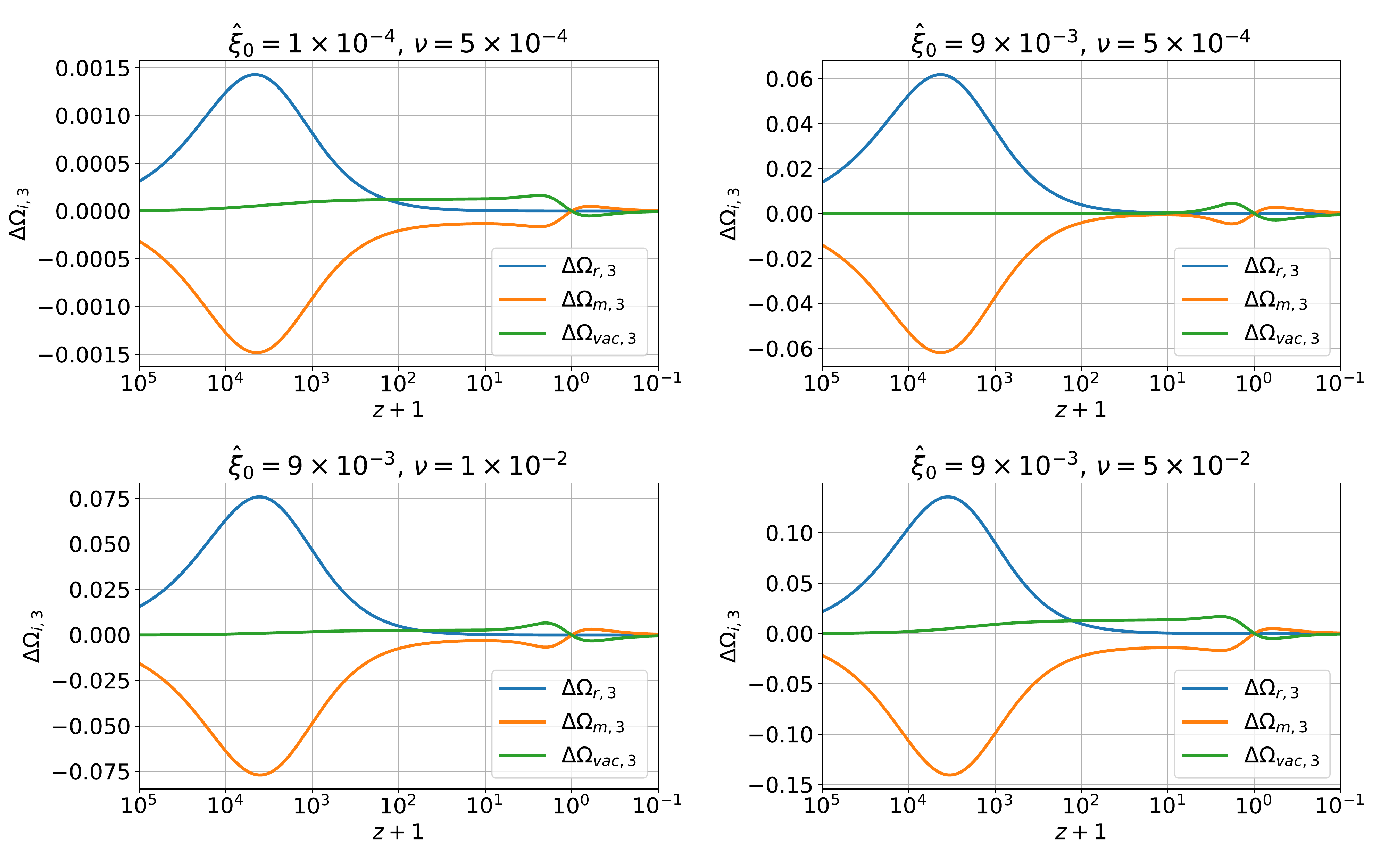}}
    \subfigure[\label{fig:M3nu-Variationdensity}]{\includegraphics[scale = 0.35]{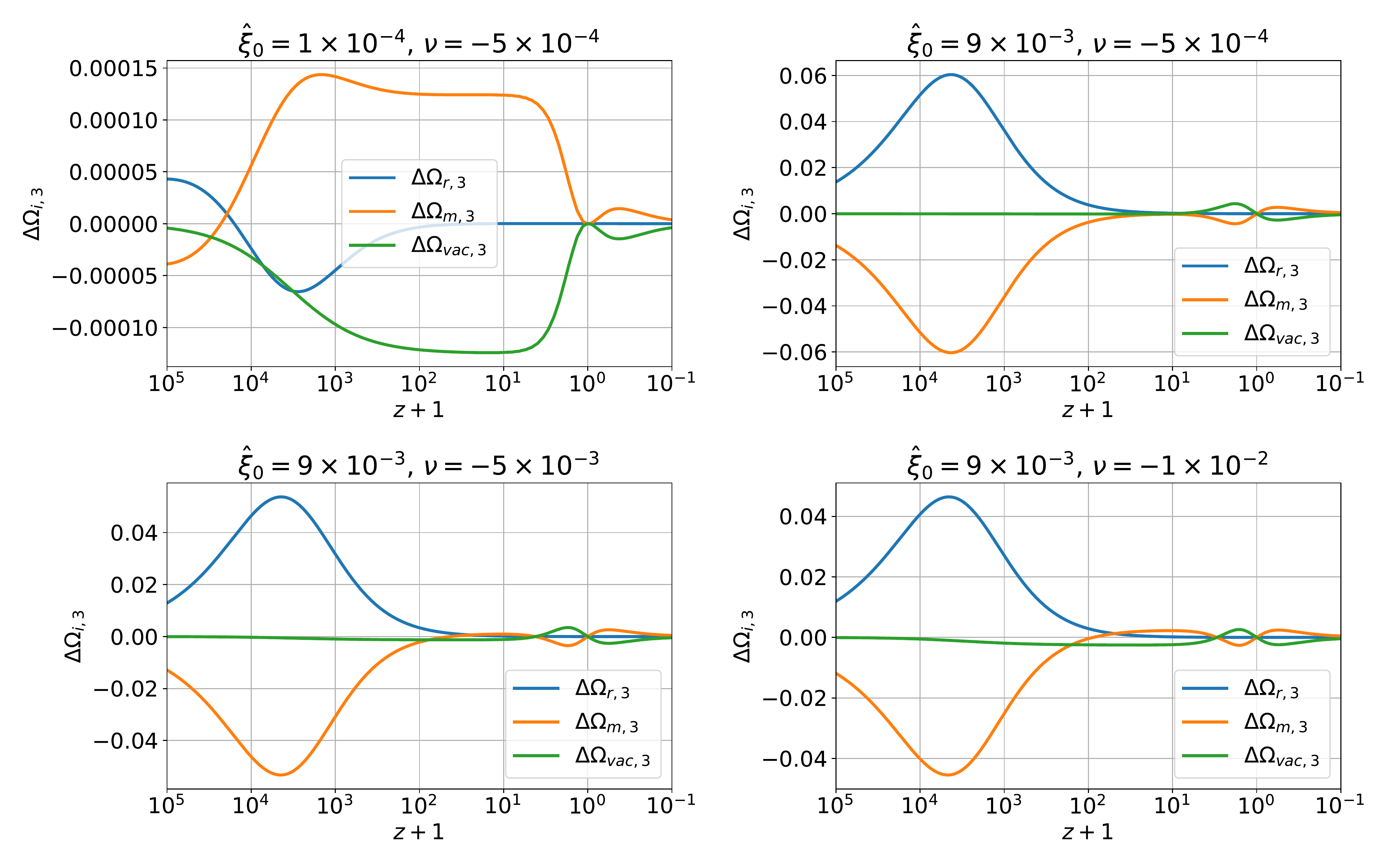}}
    \caption{\label{fig:M3VariationdensityPlots} Plots of the variation of the density parameters $\Delta\Omega_{i,3}$ associated to each fluid $\Omega_{i,3}$ for \textbf{Model 3}, with respect to their $\Lambda$CDM counterparts $\Omega_{i}$, as a function of redshift $z$, for different values of $\hat{\xi}_{0}$. Positive and negative $\nu$-values are respectively considered in \textbf{(a)} and \textbf{(b)}. The curves are obtained through the expression $\Delta\Omega_{i,3}=\Omega_{i,3}-\Omega_{i}$, where $i$ stands for $r$ (radiation), $m$ (matter), and $\rm vac$ (vacuum).}
\end{figure*}

\begin{figure*}
    \centering
    \subfigure[\label{fig:M3nu+Barotropic}]{\includegraphics[scale = 0.35]{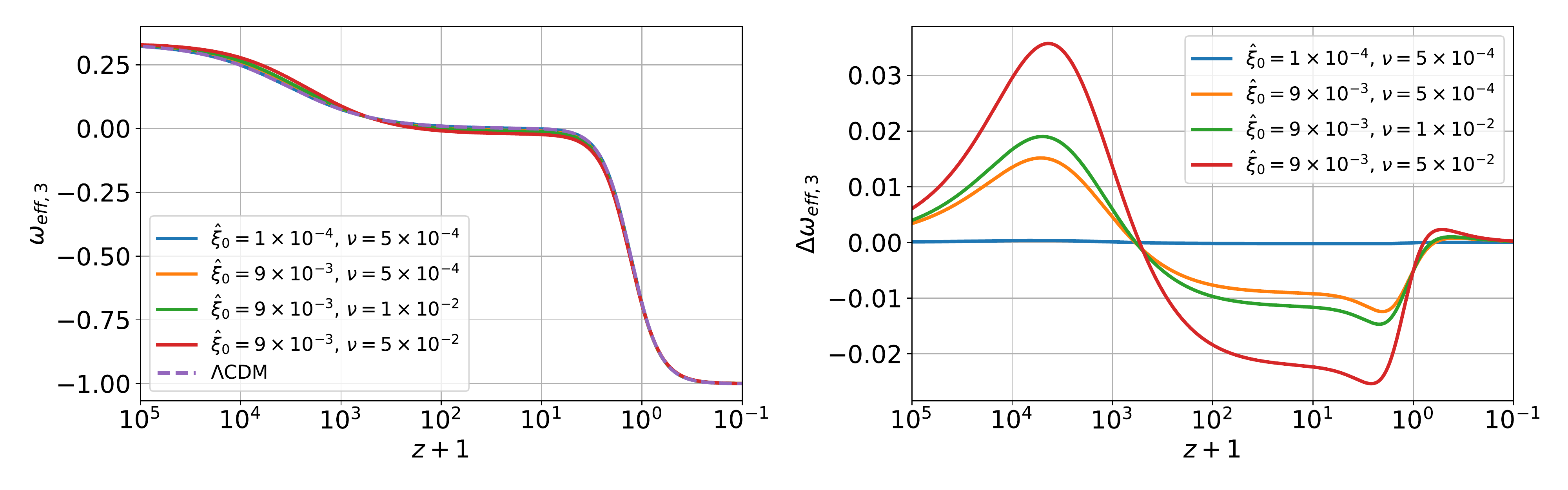}}
    \subfigure[\label{fig:M3nu-Barotropic}]{\includegraphics[scale = 0.35]{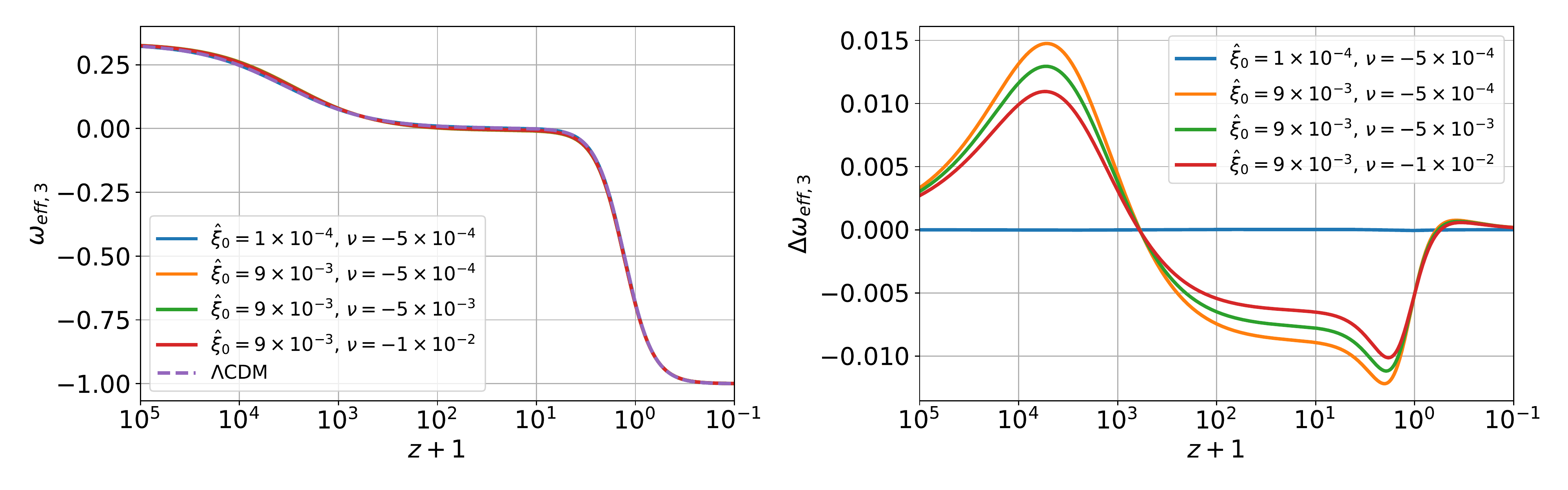}}
    \caption{\label{fig:M3BarotropicPlots} \textbf{(left)} Plot of the effective barotropic index $\omega_{\rm eff,3}$ for \textbf{Model 3}, obtained from Eq. \eqref{sec3:eqn5}, as a function of redshift $z$. \textbf{(right)}. Plot of the variation of the effective barotropic index $\Delta\omega_{\rm eff,3}=\omega_{\rm eff,3}-\omega_{\rm eff}$, were $\omega_{\rm eff}$ corresponds to their $\Lambda$CDM counterpart obtained from Eq. \eqref{sec5:eqn2}, as a function of redshift $z$. Positive and negative $\nu$-values are respectively considered in \textbf{(a)} and \textbf{(b)}, for the same values of $\hat{\xi}_{0}$.}
\end{figure*}

\begin{figure*}
    \centering
    \subfigure[\label{fig:M3nu+Vacuum}]{\includegraphics[scale = 0.35]{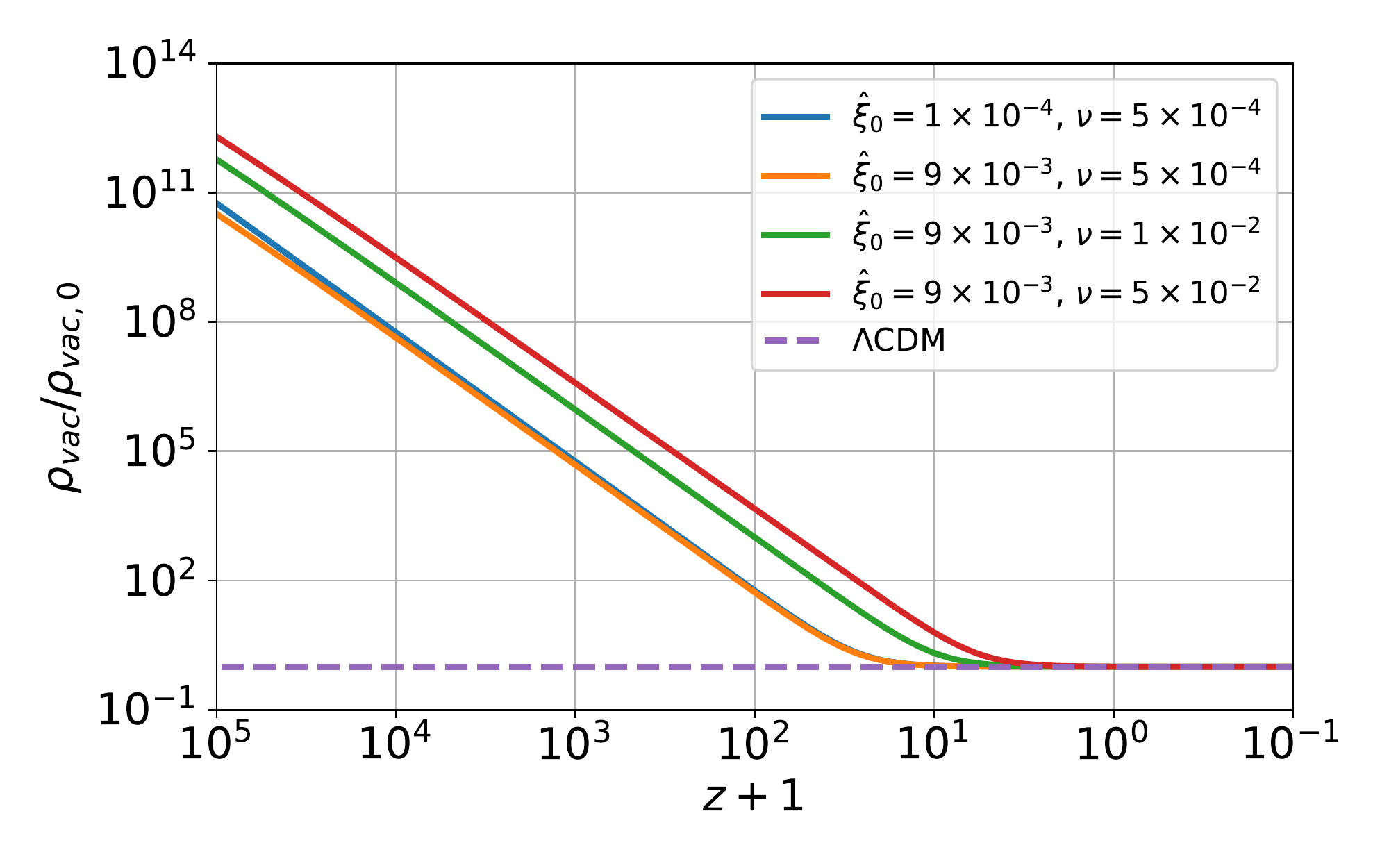}}
    \hspace{0.15cm}
    \subfigure[\label{fig:M3nu-Vacuum}]{\includegraphics[scale = 0.35]{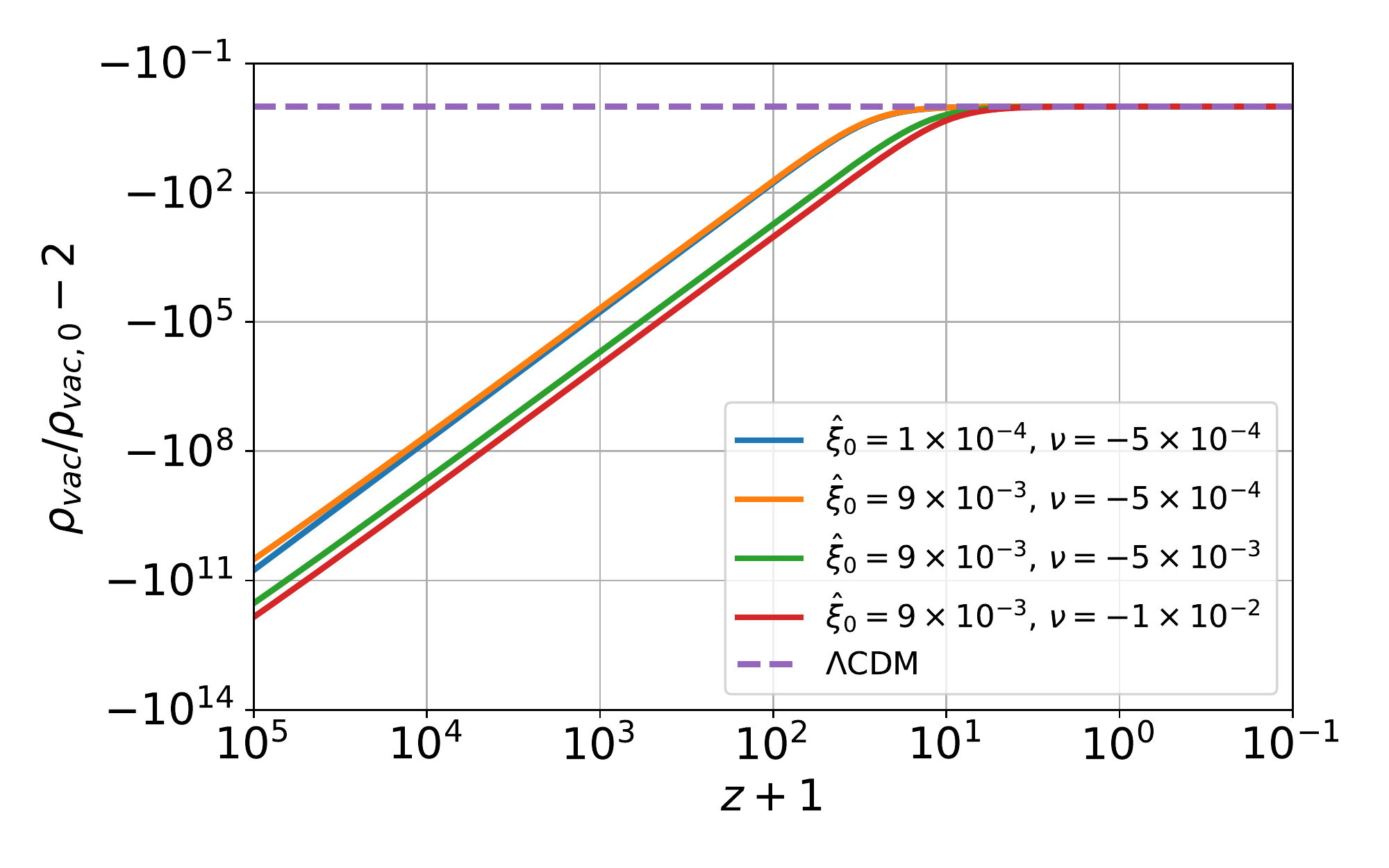}}
    \caption{\label{fig:M3Vacuum} Plots of vacuum energy density $\rho_{\rm vac}$ for \textbf{Model 3} against the redshift, normalized with respect to their current value $\rho_{\rm vac,0}$, as a function of redshift $z$. Positive and negative $\nu$-values are respectively considered in \textbf{(a)} and \textbf{(b)}, for different $\hat{\xi}_{0}$-values. Notice that we have plotted $\rho_{\rm vac}/\rho_{\rm vac,0}-2$ in order to obtain a better representation in the symmetrical logarithm scale. As a reference, we have used $\Lambda=4.24\pm0.11\times 10^{-66}$ eV$^{2}$ \cite{Planck:2018vyg} to compute the present vacuum energy density $\rho_{\rm vac,0}$.}
\end{figure*}

In figures \ref{fig:M3DensityPlots}, \ref{fig:M3VariationdensityPlots}, \ref{fig:M3BarotropicPlots}, and \ref{fig:M3Vacuum}, we present the numerical results for the model 3, obtained by the integration of Eq. \eqref{sec3:eqn4} with $s=1/2$. As a reminder, this model correspond to the second class of models where $\tilde{\nu}=\nu/2$. 

In figure \ref{fig:M3DensityPlots}, we depict the density parameters $\Omega_{i,3}$ associated to each fluid as a function of redshift $z$, as well as the density parameters $\Omega_{i}$ associated to each fluid for $\Lambda$CDM model, according to Eq. \eqref{LCDM}, for a further comparison. In figure \ref{fig:M3nu+Density} are presented the numerical results of the density parameters obtained for the different values of $\hat{\xi}_{0}$ and positive $\nu$, while in figure \ref{fig:M3nu-Density} are presented the numerical result obtained for the different values of $\hat{\xi}_{0}$ and negative $\nu$. From these figures we can see how the bulk viscosity and the running vacuum affect the redshift value in which $\Omega_{r,3}=\Omega_{m,3}$ ($z_{eq,3}$), without any appreciate effect in the redshift value in which $\Omega_{m,3}=\Omega_{\rm vac,3}$. In this sense, and similarly to what happens in the model 2, both positive and negative $\nu$ implies that $z_{eq,3}<z_{eq}$, being $z_{eq}$ the redshift range in which $\Omega_{r}=\Omega_{m}$; and the values of $z_{eq,3}$ are more closer to $z_{eq}$ for $\nu<0$ in comparison to their counterparts in the model 1. This analysis is in agreement with the dynamical system analysis made in the section \ref{systemmodel3}, where was indicated that the critical points of this model are the same of those that correspond to the model 2 (with one critical point of the model 1). Therefore considering that in the critical points exclusive of this model, presented in the table \ref{M3:critical points}, the radiation dominated period is absent as a critical point, it is an expected result that this model behaves similarly to the model 2. It is important to mention that some differences between this model and model 2 is due to the fact that, in this model, the figures are presented in the range $10^{5}\leq z+1\leq 0.1$ due to numerical difficulties.

In figure \ref{fig:M3VariationdensityPlots}, we depict the variation of the density parameters associated to each fluid with respect to the $\Lambda$CDM model as a function of redshift $z$, according to the expression $\Delta\Omega_{i,3}=\Omega_{i,3}-\Omega_{i}$. In figure \ref{fig:M3nu+Variationdensity} are presented the numerical results of the variation of the density parameters obtained for the different values of $\hat{\xi}_{0}$ and positive $\nu$, while in figure \ref{fig:M3nu-Variationdensity} are presented the numerical result obtained for the different values of $\hat{\xi}_{0}$ and negative $\nu$. From these figures we can see (with better detail than what it is seen in the figure \ref{fig:M3DensityPlots}) how the bulk viscosity and the running vacuum affect the evolution of the density parameters $\Omega_{r,3}$, $\Omega_{m,3}$, and $\Omega_{\rm vac,3}$. Following this line, and similarly to what happens in the model 2, note how a positive $\nu$ implies a greater value of $\Omega_{\rm vac,3}$ in comparison to $\Omega_{\Lambda}$; while a negative $\nu$ implies a lower value of $\Omega_{\rm vac,3}$ in comparison to $\Omega_{\Lambda}$. But, contrary to what happens in model 1, this differences occurs only at low redshift because at high redshift $\Omega_{\rm vac}$ becomes, appreciably, negligible. Unfortunately, due to the numerical issues, we are not able to ensure that at very high redshift there is not remarkably differences in the three fluids with respect to $\Lambda$CDM model as in model 2. Nevertheless, this behaviour is possible taking into account the similar behaviour with respect to model 2 seen above. Again, this is an expected result considering that the $\dot{H}$ term in the expansion for $\rho_{\rm vac}$, given by Eq. \eqref{sec2:eqn1}, represents a negative contribution while the $H^{2}$ term is a positive contribution (for $\nu<0$ the behaviour of these terms change, but, the contribution of $H^{2}$ is more important than $\dot{H}$ leading to the less contribution of $\Omega_{\rm vac, 3}$ with respect to $\Omega_{\Lambda}$). It is important to note that, when we compare these figures with their model 2 counterparts, we can see how the bulk viscosity affects the evolution of the density parameters by the election of the power $s$. This is notable for the case $\nu<0$ with $\hat{\xi}_{0}=1\times 10^{-4}$ and $\nu=-5\times 10^{-4}$.

In figure \ref{fig:M3BarotropicPlots}, we depict the effective barotropic index $\omega_{\rm eff,3}$, according to Eq. \eqref{sec3:eqn5}, and their variation with respect to the effective barotropic index $\omega_{\rm eff}$ of $\Lambda$CDM model, obtained from Eq. \eqref{sec5:eqn2}, through the expression $\Delta\omega_{\rm eff,3}=\omega_{\rm eff,3}-\omega_{\rm eff}$. In figure \ref{fig:M3nu+Barotropic} are presented the numerical results of the barotropic index and their respective variation with respect to the $\Lambda$CDM model obtained for the different values of $\hat{\xi}_{0}$ and positive $\nu$, while in figure \ref{fig:M3nu-Barotropic} are presented the numerical result obtained for the different values of $\hat{\xi}_{0}$ and negative $\nu$. From these figures we can see how the bulk viscosity and the running vacuum affect the evolution of $\omega_{\rm eff,3}$, being the most remarkably differences when $|\nu|$ takes grater values, similarly to what happens in model 2. Indeed, there is not appreciable difference between these figures and their corresponding model 2 counterparts. Therefore, as in model 2, regardless the sign of $\nu$ the values of $\omega_{\rm eff,3}$ are greater than the $\omega_{\rm eff}$ values at high redshift, with a change of this behaviour at low redshift (similar to what happens in model 1 for $\nu<0$). Unfortunately, due to the numerical issues, we are not able to ensure that at very high redshift there is not remarkably differences between $\omega_{\rm eff,3}$ and $\omega_{\rm eff}$ as in model 2. Nevertheless, this behaviour is possible taking into account the similar behaviour with respect to model 2 seen above.

In figure \ref{fig:M3Vacuum}, we depict the vacuum energy density, normalized with respect to their current value, as a function of redshift $z$, as well as the normalized vacuum energy density for $\Lambda$CDM model for a further comparison. In figure \ref{fig:M3nu+Vacuum} are presented the numerical results of the normalized vacuum energy density obtained for the different values of $\hat{\xi}_{0}$ and positive $\nu$, while in figure \ref{fig:M3nu-Vacuum} are presented the numerical result obtained for the different values of $\hat{\xi}_{0}$ and negative $\nu$. From this figures we can see how the bulk viscosity and the running vacuum affect the evolution of the vacuum energy density. In this sense, and similarly to what happens in model 2, it is clearly to see that an increment in $\hat{\xi}_{0}$ does not affect remarkably the evolution of $\rho_{\rm vac}$, contrary to what happens when we increment the values of $|\nu|$ (but note that this effect is not negligible). Again, this behaviour is due to the fact that the bulk viscosity affects the evolution of $\rho_{\rm vac}$ through the Hubble parameter and its time derivative according to the Eq. \eqref{sec2:eqn1} and, therefore, it is necessary a remarkably difference in the evolution of $H$ that does not appear due to the small values of $\hat{\xi}_{0}$. On the other hand, depending on the sign of $\nu$ it is possible to obtain an always positive vacuum energy density when $\nu>0$ or a vacuum energy density that experience a transition between a positive to a negative one when $\nu<0$. This last one occurs when the equality given by Eq. \eqref{transitiondotH} is fulfilled. Hence, we obtain an evolution of $\rho_{\rm vac}$ in this model similar to model 2, using as argument all the previous analysis made to this model (and comparing these figures with their model 2 counterparts).

\section{\label{sec:Remarks} Conclusions and final remarks}

In the present article, we have performed a detailed study of two concrete running vacuum models under a non-perturbative dynamical system perspective including, additionally, dissipation of the matter component through a general bulk viscosity given by the Ansatz $\hat{\xi}\sim H^{1-2s} \rho_{m}^{s}$, which has been recently proposed in \cite{Gomez:2022qcu}.
To be more precise, we have combined two non-trivial effects usually studied separately: 
i) a running vacuum scenario, which basically assumes that the vacuum energy density, $\rho_{\text{vac}}$, is replaced by its scale-dependent/running counterpart, enriching the potential cosmological solutions of the associated models, and 
ii) a more realistic dissipative fluid description of matter, parameterized by a non-trivial bulk viscosity, $\hat{\xi}$, which depends on the combination of DM energy density $\rho_{m}$ and on the Hubble parameter, providing a slightly but relevant modified cosmic evolution of the universe. It is worthwhile mentioning that the inclusion of bulk viscosity under a microscopic point of view is still an open task, which was not addressed here. \\
Concretely, we were mainly interested in two classes of models: 
i) $\tilde{\nu}=0$, i.e., ignoring the derivative term contained into the Ansatz for the vacuum energy density, and ii) setting $\nu=2\tilde{\nu}$ since it has the potential advantage of alleviating some tensions permeated in the $\Lambda$CDM cosmological model \cite{SolaPeracaula:2021gxi}. Notice that, albeit we parameterize our Ansatz only by $\nu$ as $\tilde{\nu}=\nu/2$, the combined effects of $H$ and $\dot{H}$ are still present and hence have non-trivial physical consequences.
Let us reinforce that Model 1 corresponds to the first class of model, given by Eqs. \eqref{sec3:eqn3}, with $s = 1/2$; while Models 2, 3, and 4 correspond to the second class of model, given by Eqs. \eqref{sec3:eqn4}, with $s=1$, $1/2$, and $0$, respectively (Table \ref{Models} gives a resume of our notations). In this respect, the take-home-message for each model is summarized as follows:
\begin{itemize}
    \item Model 1: The critical point (Ia) obviously describes a non-canonical radiation point as $\nu$ appears in its associated energy density parameter (and naturally in the effective EoS parameter $\omega_{\text{eff}}$). Also, notice that this critical point does not include the bulk viscosity parameter $\hat{\xi}_0$. By contrast, the critical point (Id) includes both, running effects and the bulk viscosity dissipation, giving rise to the same  $\omega_{\text{eff}}$ observed in case (Ia). Such ``coincidence" means that there is a sort of degeneracy, i.e., in principle, we could not distinguish the fixed points (Ia) and (Id) by reading only the equation of state, in spite of that they give rise to a different cosmological evolution.
    Point (Ie) is not altered by the inclusion of bulk viscosity and running vacuum models, and that point posses the same $\omega_{\text{eff}}$ than the critical point (Ic), i.e., one again is unable to recognize the effects of bulk viscosity (which is evident in $\rho_{\text{vac}}$) just by checking the effective EoS. 
    Point (Ib) accounts for dark matter domination and includes running effects (of the vacuum energy density) and also the effects of the bulk viscosity but on the EoS only. 
    
    Finally, Table III summarizes the corresponding Eigenvalues and stability conditions for these points. Irrespective of the precise values of critical points, the impact of running vacuum appears to be more profound in setting the stability conditions than the modifications coming from bulk viscosity. Also, the phase space diagram reveals that the system shows an attractor-character after a saddle-like behaviour, independent of the initial conditions, but strongly dependent of the model parameters. For instance, if $\hat{\xi}_0\sim \mathcal{O}(1)$ the DM-like fixed point corresponds to an attractor, while for $\hat{\xi}_0\ll1$, it corresponds to a saddle point. The former case represents, indeed, an unified dark fluid scenario, in which the acceleration expansion is driven by the dark matter component, and the latter corresponds to the usual cosmic evolution, being the DE responsible for the universe acceleration expansion. 
    \item Model 2: This second case accounts for the more general situation addressed in this paper, i.e., $\tilde{\nu} \neq 0$. Here we have four critical points, remarkably two of them (IIa and IIb) are not susceptible to the inclusion of running of the vacuum energy density and dissipation, as the density parameters for these points are independent of the values of $\nu$ and $\hat{\xi}_0$. On the contrary, points IIc and IId are strongly dependent of $\nu$ and $\hat{\xi}_0$. In particular, notice that point IIc encodes an intermediate period which is categorized as non-standard dark matter, involving simultaneously the effects due to the running and the viscosity. 
    Interestingly, for a weak $\nu$-coupling the energy density parameters associated to matter and vacuum energy density are slightly modified as follows:
    \begin{align}
        \Omega_m & = 1 - \frac{1}{4} \nu  \Bigl(1 + 3 \hat{\xi}_0 \Bigl) \ + \ \mathcal{O}(\nu^2),
        \\
        \Omega_{\text{vac}} & = \frac{1}{4} \nu  \Bigl(1 + 3 \hat{\xi}_0 \Bigl) \ + \ \mathcal{O}(\nu^2).
    \end{align}
    Point IId is even more involved, corresponding to standard radiation domination for a concrete value of $\nu$, but in general, it corresponds to DE domination. 
    With respect to the stability of this model we can confirm, for certain values of $\nu$ and $\hat{\xi}_0$ that:
    i) point IIa is a repeller
    ii) point IIb is an attractor
    iii) point IIc could be saddle or attractor, and
    iv) point IId is an attractor.
    \item Model 3: This case is based on the differential system Eq.~(\ref{sec3:eqn4}) with $s=1/2$. Most of the properties of this model also appear in Model 1 and Model 2, hence, we will not discuss them again. Instead, we will comment the novelty: taking the Ansatz Eq. (\ref{sec2:eqn5}) we notice the existence of a (viscosity-running) two-parameters family of solutions. In fact, points (IIc) and (IIIa) have the same effective EoS in the $\hat{\xi}_0 \to 0$ limit.
    \item Model 4: 
    Consistent with the result achieved in our previous paper \cite{Gomez:2022qcu}, in which the running effects were not included, the present model with exponent $s=0$ must be discarded as it can not describe successfully the whole cosmological evolution of the universe. Unfortunately, the inclusion of a running vacuum energy density does not modify such fact, i.e., in the present case the 
    radiation-dominated period is still absent, the reason why this cosmological scenario is not of physical interest.
\end{itemize}
Thus, the inclusion of both, a running vacuum energy density and a dissipative DM component enrich the whole physical scenario by adding non-trivial critical points absent in their non-running and non-viscous counterparts. 
For example, the inclusion of bulk viscosity can drive the current acceleration expansion of the Universe and provide a new phantom-like behaviour, depending on the values of $\hat{\xi}_0$ and $\nu$. We notice that the effect of the running vacuum energy density is not present in the late-times dynamics of the universe for the realization of the model 1.

The results obtained from the dynamical system analysis were complemented by performing the numerical integration of the models 1, 2, and 3, discarding beforehand the model 4 due to the absence of a dominant radiation critical point.
In this sense, the numerical results for the successful models show the capability of describing three eras of the cosmic evolution: radiation, matter and dark energy domination eras.

An interesting feature is that the corresponding redshift value ($z_{eq,i}$) at which $\Omega_{r,i}=\Omega_{m,i}$ (where $i$ stands for models 1, 2, and 3) depends on the particular model; while the redshift value at which $\Omega_{m,i}=\Omega_{\text{vac},i}$ remains slightly unchanged compared with the prediction of $\Lambda$CDM for all these three models. This is a very desirable result for the proposed extension of the standard cosmological model, in agreement with some observational inferences.
Nevertheless, the model 1 has the feature that a non-trivial combination of values $\nu$ and $\hat{\xi}_0$ may lead to the same redshift value at which $\Omega_{r,1}=\Omega_{m,1}$, i.e., $z_{eq,1}=z_{eq}$, while for the other models we were unable to find suitable values to fulfill this property. 

On the other hand, all these models lead to a larger contribution of $\Omega_{\rm vac}$ when compared to the $\Lambda$CDM model, for positive $\nu$-values. On the contrary, negative values of $\nu$ reduce $\Omega_{\rm vac}$.
Interestingly, the difference of the vacuum density parameter, $\Delta \Omega_{\rm vac}$, presents a nearly constant variation for the model 1 from $z+1 \approx 10$ up to very high redshift, which appreciably deviates from the value of the standard cosmological model, as it can be seen in fig. \ref{fig:M1VariationdensityPlots}. Nevertheless, in models 2 and 3 such differences become approximately neglectable at very high redshift. This feature comes from the extra term present in $\rho_{\rm vac}$, which depends on $\dot{H}$ and gives a negative contribution.
The same reasoning applies to the behavior of $\Delta \omega_{\rm eff}$ for these models. In fact, as the redshift increases this difference goes to zero, as it can be seen in fig. \ref{fig:M3BarotropicPlots}. 

%

Finally, when compared the vacuum energy density at high redshift to its observed current value, it turns out that the largest difference occurs for model 1, which is again due to the contribution of $\dot{H}$ to $\rho_{\rm vac}$ in these models.
A remarkable feature of $\rho_{\rm vac}$ is its sign change during the matter domination period depending on the sign of $\nu$. This change of sign occurs according to Eq. \eqref{transitionnotdotH} for model 1 and Eq. \eqref{transitiondotH} for models 2 and 3. 
It is important to mention that the leading contribution to the variation of $\rho_{\rm vac}$ comes from $\nu$. Nevertheless, the dissipative effects certainly affect the whole evolution of the system as was inferred from the dynamical system analysis.

 As an overall and encouraging conclusion from the dynamical system analysis is that we have obtained new critical points of cosmological relevance that arise from the combined effects of a running vacuum energy density and dissipative DM component. Hence, the associated cosmological models are characterized by two parameters describing the above-mentioned effects. This is appealing to the light of current tensions permeated in the $\Lambda$CDM model, because of the potentiality to successfully describe the current observations due to an enlarged parameter phase space characterizing these extended models.
Further constraints on the model parameters can be obtained by means of high-quality observational constraints by using the derived parameter regions from the dynamical system analysis as priors for the statistical treatment. This task will be addressed in future works.
\section*{Acknowledgements}

G. G. acknowledges the financial support from Agencia Nacional de Investigaci\'on y Desarrollo (ANID) through the FONDECYT postdoctoral Grant No. 3210417. E. G. acknowledges the support of Direcci\'on de Investigaci\'on y Postgrado at Universidad de Aconcagua. G. P. acknowledges the financial support from Dicyt-USACH Grant No. 042231PA.
A.R. and N.C. acknowledge Universidad de Santiago de Chile for financial support through the Proyecto POSTDOCDICYT, C\'odigo 042231CM-Postdoc.

\bibliographystyle{apsrev}
\bibliography{sample2}

\end{document}